%% file: article_and_supplemental.tex
\documentclass[twocolumn,secnumarabic,amssymb,nobibnotes,aps,showpacs,noeprint,superscriptaddress]{revtex4-1}
\usepackage[english]{babel}
\usepackage{graphicx}

\usepackage{amsmath}
\usepackage{amsfonts}
\usepackage[usenames,dvipsnames]{color}

\usepackage{bm}
\usepackage{natbib}

\usepackage{hyperref}
\usepackage{rotating}
\usepackage{booktabs}
\usepackage{slashed}
\usepackage{transparent}

\usepackage{color}

\usepackage{nameref}

\usepackage[utf8]{inputenc}
\usepackage[T1]{fontenc}

\usepackage{cleveref}

\usepackage{amssymb}

\usepackage{longtable}

\usepackage{cleveref}

\makeatletter
\def\maketitle{
\@author@finish
\title@column\titleblock@produce
\suppressfloats[t]}
\makeatother

\begin{document}

\title{Anomalous Underscreening in the Restricted Primitive Model}

\author{Andreas H\"artel}
\email{andreas.haertel@physik.uni-freiburg.de}
\affiliation{Institute of Physics, University of Freiburg, Hermann-Herder-Stra{\ss}e 3, 79104 Freiburg, Germany}

\author{Moritz B\"ultmann}
\affiliation{Institute of Physics, University of Freiburg, Hermann-Herder-Stra{\ss}e 3, 79104 Freiburg, Germany}

\author{Fabian Coupette}
\affiliation{Institute of Physics, University of Freiburg, Hermann-Herder-Stra{\ss}e 3, 79104 Freiburg, Germany}

\date[]{published in: Physical Review Letters \textbf{130}, 108202 (2023), DOI: \href{https://doi.org/10.1103/PhysRevLett.130.108202}{10.1103/PhysRevLett.130.108202}}

\begin{abstract}
Underscreening is a collective term for charge correlations in electrolytes decaying slower 
than the Debye length. Anomalous underscreening refers to phenomenology that cannot be attributed 
alone to steric interactions. Experiments with concentrated electrolytes and ionic fluids report 
anomalous underscreening, which so far has not been observed in simulation. We present Molecular 
Dynamics simulation results exhibiting anomalous underscreening that can be connected to cluster 
formation. A theory that accounts for ion pairing confirms the trend. 
Our results challenge the classic understanding of dense electrolytes 
impacting the design of technologies for energy storage and conversion. 
\end{abstract}

\maketitle

In recent years, 
unexpectedly long decay lengths of electrostatic forces have been observed in concentrated 
electrolytes \cite{gebbie_pnas110_2013,gebbie_pnas112_2015,cheng_ami2_2015,
espinosa-marzal_jpcl5_2014,cheng_ami2_2015,
smith_jpcl7_2016,hjalmarsson_cc53_2017,smith_jpcb123_2019,gaddam_lang35_2019} 
subsumed under the term ``underscreening''. 
A lot of effort has been committed to explaining underscreening \cite{gebbie_cc53_2017,
lee_prl119_2017,lee_fd199_2017,goodwin_ec82_2017,kjellander_jcp148_2018,rotenberg_jpcm30_2018,
coupette_prl121_2018,adar_pre100_2019,coles_jpcb124_2020,zeman_cc56_2020,anousheh_aipa10_2020,
kjellander_pccp22_2020,zeman_jcp155_2021,ciach_jpcm33_2021,krucker-velasquez_jcp155_2021,
cats_jcp154_2021,kumar_jcis622_2022}. We distinguish regular underscreening that can 
be attributed to steric interactions from anomalous underscreening characterized by much 
longer decay lengths compared to its regular counterpart. 
Numerous studies have concluded that one of the most fundamental models for electrolytes 
and ionic liquids, 
the restricted primitive model (RPM), does not exhibit anomalous underscreening.
As even some experimental studies could not find these large decay lengths 
\cite{schoen_bjn9_2018,kumar_jcis622_2022}, the phenomenon itself has been questioned. 
In this Letter, we demonstrate that there is anomalous underscreening in the RPM using 
Molecular Dynamics simulations.
However, our findings do not support a unique scaling of decay 
lengths as reported in \cite{lee_prl119_2017,lee_fd199_2017}. 
We can explain our directly measured results with cluster formation, which effectively reduces the 
concentration of mobile charge carriers. Finally, we propose a minimal theory 
of ion pairing 
that captures the phenomenology and even provides sensible agreement with the experiment.

In an ionic fluid, the Coulomb interaction between two charged particles is 
exponentially screened due to the presence of mobile charge carriers. 
The screening length is the inverse decay rate of this exponential, which 
reflects the ability of an electrolyte to screen surface 
charges on electrodes. Accordingly, it is closely related to the formation 
of electric double layers, which play a fundamental role in, among 
others, modern charge storing, energy conversion, and desalination technologies 
\cite{simon_nm7_2008,brogioli_prl103_2009,haertel_ees8_2015,porada_pms58_2013}, 
chemical and colloidal interactions \cite{verwey_book_1948,derjaguin_pss43_1993,li_pnas114_2017}, 
and DNA \cite{kornyshev_rmp79_2007,espinosa_s19_2019},
as well as nervous conduction \cite{shapiro_nc3_2012,lichtervelde_pre101_2020}. 
The strength of electrostatic interactions is encoded in the Bjerrum length 
$\lambda_{\textrm{B}}=e^2/(4\pi\epsilon_{0}\epsilon k_{\textrm{B}} T)$, with 
elementary charge $e$, 
vacuum permittivity $\epsilon_{0}$, the relative dielectric 
permittivity of the solvent $\epsilon$, and Boltzmann's constant $k_{\textrm{B}}$. 

The expected decay length for dilute systems of charged particles 
is given by the Debye screening length 
$\lambda_{\mathrm{D}}=1/\sqrt{8\pi \rho_{\textrm{s}} \lambda_{\mathrm{B}}}$ \cite{debye_pz24_1923} 
that decreases with increasing number density $\rho_{\textrm{s}}$ of mobile 
charges and with the Bjerrum length. By convention $\rho_{\textrm{s}}$ is the 
individual density of positive and negative charges, respectively, and 
often given as salt concentration $c$. 
Underscreening refers to a less effective screening, i.e., decay 
lengths exceeding the Debye length that have been observed by surface force 
apparatus (SFA) experiments for high salt concentrations or large Bjerrum 
lengths \cite{gebbie_pnas110_2013}. 

However, the experiments report that the charge correlation is the sum of 
two qualitatively different decays: 
a potentially oscillatory structural decay at small distances and 
a much slower long-ranged strictly nonoscillatory decay at greater 
separations \cite{smith_prl118_2017}. 
The structural decay is well understood theoretically within the RPM of 
charged hard spheres and originates from the interplay between electrostatic 
and steric interactions of the particles (see \cite{cats_jcp154_2021} 
and references therein).

From simulations of the monovalent RPM, we can extract the charge correlation 
as $h_{\textrm{cc}}=g_{++}-g_{+-}$, where $g_{\mu \nu}$ denotes the species 
resolved pair-distribution function. 
In theory, these pair-distribution functions can be obtained from the 
Ornstein-Zernike equation that defines their analytic structure. 
The charge correlation can be expressed as an infinite sum over terms 
of the form
\begin{align}
H_i(r) = A_i \exp(-r/\lambda_i)\cos(\omega_i r+\tau_i)/r^{b_i} 
\label{eq:pole-correlation-function}
\end{align}
with decay or screening length $\lambda_i$, amplitude $A_i$, 
$\omega_i$ and $\tau_i$ describing potential oscillations, 
and $b_i \in \{1,2\}$. 
Each term originates from a complex singularity of an auxiliary function, 
$b_i = 1$ applies for simple poles \cite{attard_pra45_1992,
kjellander_jpl200_1992,evans_mp80_1993,evans_jcp100_1994}, and $b_i=2$ for branch points 
\cite{ennis_jcp102_1995,ulander_jcp114_2001}. 
Further details are given in the Supplemental Material 
\footnote{See Supplemental Material at \url{http://link.aps.org/supplemental/10.1103/PhysRevLett.130.108202} for details on the simulations, the fitting of decay lengths, and the derivation of the minimal cluster theory, which includes \cite{hansen_book_2013,weik_epjst227_2019,andersen_pra4_1971,weeks_jcp54_1971,hockney_book_1988,gonzalez-mozuelos_jcp139_2013,evans_ap28_1979,haertel_jpcm29_2017,roth_jpcm28_2016,rosenfeld_prl63_1989,roth_jpcm22_2010,hansen-goos_jpcm18_2006,percus_prl8_1962,frisch_book_1964}.}.\makebox[0pt][l]{ \phantom{\cite{hansen_book_2013,weik_epjst227_2019,andersen_pra4_1971,weeks_jcp54_1971,hockney_book_1988,gonzalez-mozuelos_jcp139_2013,evans_ap28_1979,haertel_jpcm29_2017,roth_jpcm28_2016,rosenfeld_prl63_1989,roth_jpcm22_2010,hansen-goos_jpcm18_2006,percus_prl8_1962,frisch_book_1964}} }
At long separations $r$ the contribution with the longest decay length 
$\lambda_i$ dominates. 
With increasing salt concentration, the dominant exponential decay switches 
from monotonic to oscillatory (Kirkwood crossover) 
\cite{kirkwood_jcp7_1939,leoteDeCarvalho_mp83_1994} 
as well as from charge to density dominated 
\cite{coupette_prl121_2018}, 
depending on the ionic diameter. 

It needs to be emphasized that the structural decay observed in the experiments 
already shows underscreening. This \emph{regular} underscreening has been 
observed in simulations and the underlying mechanism is theoretically well 
understood.

In contrast to that, the long-ranged decay was found exclusively in a few 
experimental studies. 
Recent works concluded that the RPM that accurately explains the structural 
decay is incapable of predicting the long-ranged decay 
\cite{zeman_cc56_2020,cats_jcp154_2021,zeman_jcp155_2021,kumar_jcis622_2022}, 
which we refer to as \emph{anomalous} underscreening. 
Thus, either the RPM is missing a crucial ingredient or the long-ranged decay 
is an artifact of the experiment. 
Within this Letter we show that there is a third option.

\begin{figure}[t]
\centering
\includegraphics[width=\linewidth]{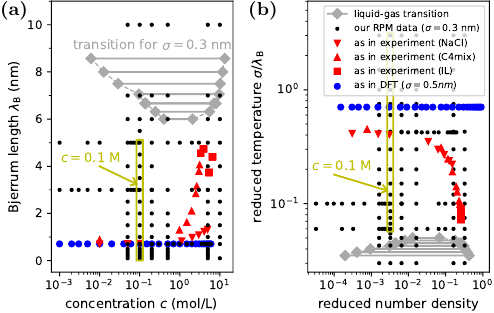}
\caption{Phase diagram of the RPM for (a) the concentration $c$ and 
Bjerrum length $\lambda_{\textrm{B}}$ 
and (b) for reduced temperature $T^{*}=\sigma/\lambda_{\textrm{B}}$ and 
total number density $\rho^{*}=2\rho_{\textrm{s}}\sigma^3$. 
Each symbol marks a parameter set for which we have run MD simulations. 
The data points for $c=0.1$ mol/L and $\lambda_{\mathrm{B}}=0.1\dots 5.0$ nm 
are highlighted by an orange rectangle.
Special symbols show the sets used 
in a previous theoretical study (DFT) \cite{cats_jcp154_2021} 
and in experiments \cite{smith_jpcl7_2016} with 
NaCl in water (NaCl), 
[C4C1Pyrr][NTf2] in propylene carbonate (C4mix), 
and an ionic liquid (IL) 
(further details in the Supplemental Material \cite{Note1}). 
Horizontal lines mark the region of liquid-gas phase coexistence 
\cite{orkoulas_jcp101_1994}; see \cite{hynninen_prl96_2006} for further phases. 
}
\label{fig:phase-diagram}
\end{figure}

Underscreening is often categorized by power laws of the form 
$\lambda/\lambda_{\textrm{D}} \sim (\sigma/\lambda_{\textrm{D}})^p$, 
with ion diameter $\sigma$, even though the available data do not cover a 
single decade. 
Regular underscreening corresponds to $p\approx 3/2$ while $p \approx 3$ 
is anomalous.
The SFA results suggest 
that $\lambda/\lambda_{\textrm{D}}$ depends uniquely on the dimensionless 
quantity $\kappa=\sigma/\lambda_{\textrm{D}}$, 
because data for many different electrolytes and ionic liquids all collapse 
onto one unique curve \cite{lee_prl119_2017}. 
This conclusion, however, is misleading. Figure~\ref{fig:phase-diagram} 
illustrates the phase diagram of the RPM in (a) dimensional and 
(b) reduced dimensionless units. 
Every small circle marks a parameter set $(c,\lambda_{\textrm{B}})$ 
for which we performed Molecular Dynamics (MD) simulations. 
The triangles and squares correspond to parameters as used in the SFA experiments 
for different electrolytes and ionic liquids. 
Curiously, in reduced units all experimentally probed parameter sets collapse 
onto one curve in the phase diagram. 
Thus, it is not surprising that the resulting decay lengths do the same. 

\begin{figure}[t]
\centering
\includegraphics[width=\linewidth]{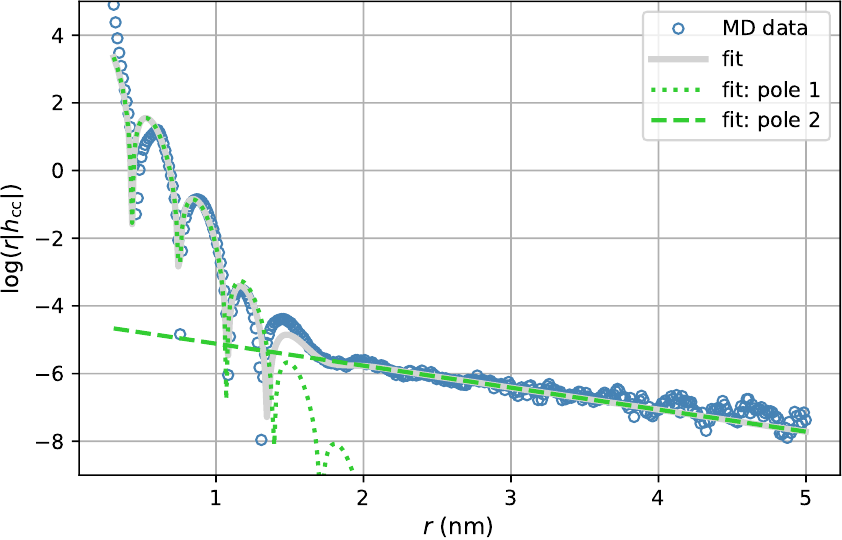}
\caption{Charge-correlation function $h_{\textrm{cc}}(r)$ in 
a representation that shows the decay length as the slope of the graph. 
This function was sampled by a MD simulation with $c=0.1$ mol/L, 
$\lambda_{\textrm{B}}=5$ nm, and $\sigma=0.3$ nm. 
The pole fits have the analytical form of \cref{eq:pole-correlation-function}, 
respectively (see Supplemental Material \cite{Note1} for further details on the fits). }
\label{fig:correlation-fit}
\end{figure}

Simulation and theoretical studies typically explore underscreening by solely 
varying the concentration exemplified by the large circles 
in \cref{fig:phase-diagram}.
Conversely, the experimental parameters that exhibit anomalous underscreening 
at large concentrations predominantly vary in the Bjerrum length. 
Thus, previous studies only explore limited parts of parameter space. 
To address this issue, we present MD simulations for a wide range of parameters 
comprehensively screening the phase diagram as illustrated 
in \cref{fig:phase-diagram}. 
In particular, this allows us to extract the decay length as a function of 
the Bjerrum length for several fixed concentrations. 
For each set of parameters $(c,\lambda_{\textrm{B}})$ we run 
MD simulations of the RPM with $\sigma=0.3$ nm [further details in the Supplemental Material \cite{Note1}; 
the typical size of the cubic simulation box is $(60$ nm$)^3$]. 
Once equilibrated, we sample the radial pair-distribution functions 
$g_{\mu\nu}(r)$ and compute the charge correlation $h_{\textrm{cc}}$. 
To extract the principal decay lengths, we fit $h_{\textrm{cc}}$ to a superposition of 
decays $H_i$ [\cref{eq:pole-correlation-function}] accounting for up to 
three poles and a potential branch point -- \cref{fig:correlation-fit} 
exemplifies the procedure. 

The representation $\log(r|h_{\textrm{cc}}|)$ is chosen in accordance with the 
known form of the decay in \cref{eq:pole-correlation-function} so that the 
decay length corresponds to the slope of a linear fit. 
In \cref{fig:correlation-fit}, we find the two previously discussed decay 
regimes: the structural decay (up to $r\approx 1.5$ nm) and long-ranged decay 
($r\gtrapprox 1.5$ nm). 
Consistent with the SFA measurements, the long-ranged decay (pole 2) is always found 
to be monotonic, while the structural decay (pole 1) can also show oscillations depending 
on the parameters. 
At very large separations, the decay with the largest decay length dominates. 
However, in the Supplemental Material \cite{Note1} we demonstrate that the amplitude $A_i$ of 
this dominant contribution may be small such that the signal is buried in 
statistical noise of the simulations.
This complicates the extraction of decay lengths, particularly for large 
concentrations.
Details on the simulations and fitting procedure including the fitted 
parameters for all charge correlations can be found in \cite{Note1}.

\begin{figure}[t]
\centering
\includegraphics[width=\linewidth]{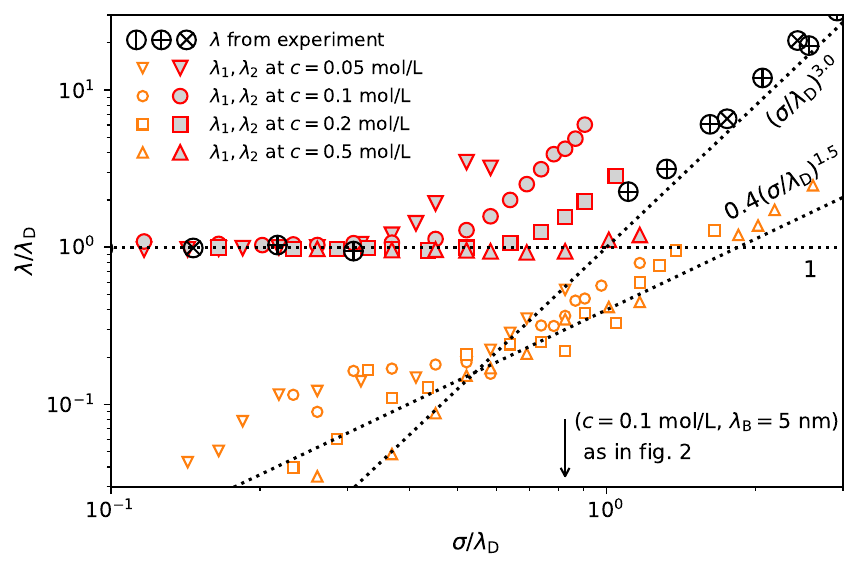}
\caption{ 
Decay lengths $\lambda_1$ and $\lambda_2$ obtained by fitting 
$\sum_{i=1}^n H_i(r)$, $n\in\{1,2,3\}$ to bulk charge-correlation functions sampled 
from our MD simulations of the RPM as exemplarily shown in \cref{fig:correlation-fit}. 
$\lambda_1$ represents decay lengths of the structural decay and $\lambda_2$ represents decay 
lengths of the long-ranged monotonic decay (compare poles 1 and 2 in \cref{fig:correlation-fit}). 
Note that in some cases we used a third pole for the fit \cite{Note1}.
We show the decay 
length $\lambda$ in relation to the Debye length $\lambda_{\textrm{D}}$ 
against $\sigma/\lambda_{\textrm{D}}$ (depending on $\lambda_{\textrm{B}}$ and $c$), 
as common in the literature on underscreening \cite{lee_fd199_2017}. 
For each given concentration, we varied only the Bjerrum length. 
Large black circles represent data from experiments on an ionic liquid (|), 
NaCl in water (+), and [C4C1Pyrr][NTf2] in propylene carbonate (X) \cite{smith_jpcl7_2016}. 
Dotted lines depict power laws as noted. 
}
\label{fig:decay}
\end{figure}

\begin{figure}[t]
\centering
\includegraphics[width=\linewidth]{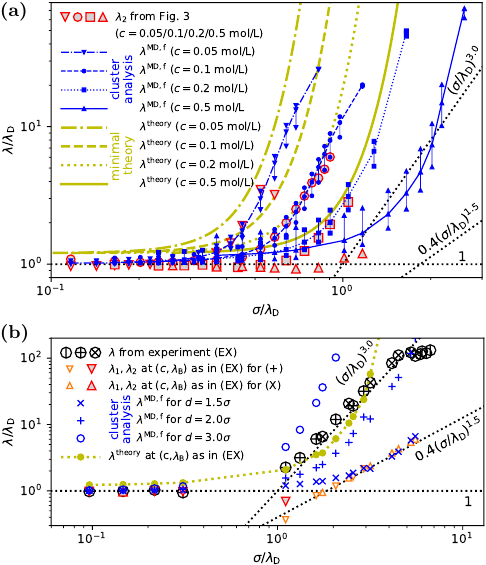}
\caption{
Decay lengths represented as in \cref{fig:decay}. 
(a) Symbols without lines show decay lengths of anomalous underscreening ($\lambda_2$ in 
\cref{fig:decay}) as obtained by fitting the charge-correlation functions from our MD simulations of the RPM. 
Triplets of vertically arranged blue symbols show 
decay lengths $\lambda^{\mathrm{MD},\mathrm{f}}$ induced by free ions 
for different 
connectivity lengths $d=1.5\sigma$, $d=2\sigma$, $d=3\sigma$ (from top to bottom) 
in the cluster search algorithm. 
Yellow lines show the prediction $\lambda^{\mathrm{theory}}$ of our minimal theory. 
(b) Directly measured $\lambda_1$ and $\lambda_2$ obtained by fitting the charge-correlation 
functions sampled from our MD simulations for parameter pairs $(c,\lambda_{\mathrm{B}})$ as 
used in the experiments (EX) of \cite{smith_jpcl7_2016}. The experimental data are described in 
\cref{fig:decay} and also listed in the Supplemental Material \cite{Note1}. 
Blue symbols show the resulting $\lambda^{\mathrm{MD},\mathrm{f}}$ from our cluster analysis 
for the same parameters. Yellow symbols (line added for clarity) show the corresponding 
$\lambda^{\mathrm{theory}}$ from our minimal theory. 
}
\label{fig:cluster-decay}
\end{figure}

In \cref{fig:decay}, we present the measured decay lengths that we obtained 
from our MD simulations at fixed concentration $c=0.05/0.1/0.2/0.5$ mol/L, 
respectively (results for all parameter pairs presented in \cref{fig:phase-diagram} 
are shown in the Supplemental Material \cite{Note1}). 
Our broad exploration of the phase diagram reveals that there is no unique 
relationship in reduced parameters. 
The decay length generally depends on salt concentration and Bjerrum length, 
independently. 
If we increase the Bjerrum length at fixed concentration, we find $\lambda_2$ 
being the Debye length at small $\sigma/\lambda_{\mathrm{D}}$ but approximately 
following a power law $\lambda_2\sim(\sigma/\lambda_{\mathrm{D}})^3$ at larger 
$\sigma/\lambda_{\mathrm{D}}$, as observed in the experiments for dense electrolytes. 
Each $\lambda_2$ is accompanied 
by a much shorter decay length $\lambda_1$ that describes structural screening. 
All structural decay lengths $\lambda_1$ approximately follow a power law 
$\lambda_1\sim(\sigma/\lambda_{\mathrm{D}})^{1.5}$, 
as demonstrated in \cref{fig:decay}. 

The curves of $\lambda_2$ at fixed concentration that show anomalous underscreening 
shift to the right in \cref{fig:decay} with increasing concentration. 
At the same time, the amplitude of the decay $H_2$ from \cref{eq:pole-correlation-function} 
decreases with increasing concentration. At high concentrations, the decay $H_2(r)$ 
drops below the numerical resolution of our MD simulation (see \cite{Note1} for further details). 
For reference, 
we also show experimental results for different ionic liquids 
and electrolytes in \cref{fig:decay}. 
Varying the concentration, 
we only find regular underscreening in the dense regime (data not shown). 
The extracted decay lengths approximately follow 
the power law $0.4(\sigma/\lambda_{\textrm{D}})^{1.5}$ (\cref{fig:decay}), 
which is consistent with the literature \cite{cats_jcp154_2021}. 

In conclusion, 
there is anomalous underscreening in the RPM but it cannot be 
observed in simulations for parameters suggested by the experiment.

However, if there is anomalous underscreening in the RPM, theoretical approaches 
should find it as well. Recently, Cats \textit{et al.} presented a comprehensive 
comparison between available theoretical results and concluded that classical 
density functional theory (DFT) is  
a good approach to describe screening in electrolytes and 
ionic liquids \cite{cats_jcp154_2021}. 
Classical DFT accurately predicts the structural decay, i.e., 
regular underscreening 
\cite{coupette_prl121_2018,cats_jcp154_2021,bueltmann_jpcm24_2022,kumar_jcis622_2022}. 
However, DFT calculations for a fixed concentration $c=0.1$ mol/L and varying 
Bjerrum length do not show anomalous underscreening \cite{Note1}, 
in contrast to our MD simulations (\cref{fig:decay}). 
The predictions of classical DFT reflect the accuracy of the employed excess 
free energy functional. It stands to reason that the theory does simply not 
account for the mechanism that causes anomalous underscreening.

Candidates for missing ingredients in the theoretical 
description are the subject of ongoing discussions. 
Theoretical models are frequently criticized for their implicit treatment of solvents 
that can significantly alter the effective steric and electrostatic interactions between ions. 
However, anomalous underscreening has been reported experimentally for a variety of very 
different solvents and even for ionic liquids. 
Moreover, simulations that explicitly accounted for atomistic solvent did not observe anomalous 
underscreening \cite{zeman_cc56_2020}. 
A promising contender is a reduction of the concentration of effective charge 
carriers, for instance, by the formation of Bjerrum pairs or by defects in 
dense electrolytes taking over the role of mobile charges 
\cite{zwanikken_jpcm21_2009,gebbie_pnas110_2013,richey_jacs135_2013,
gebbie_pnas110response_2013,adar_jcp146_2017,feng_prx9_2019,smith_jpcb123_2019,
jones_jcp154_2021,dean_prl127_2021,krucker-velasquez_jcp155_2021,goodwin_jcp157_2022,
jones_arxiv_2022}. 
To estimate the effective concentration of free charge carriers, 
we analyze system configurations generated by our MD simulations for cluster 
formation. 

To this end, we assign a connectivity shell of diameter $d>\sigma$ to all 
particles in our simulation and consider two particles connected if their 
respective connectivity shells overlap. 
The clusters detected in this way 
either comprise the same number of positive and negative charges, such 
that their collective contribution to screening is supposedly negligible, 
or have a finite net charge. Based on our cluster results, we safely assume that 
the absolute net charge of a cluster is either one elementary charge or zero \cite{Note1}. 
In consequence, we define free ions by neglecting all clusters that contain 
an even number of particles and by replacing each cluster that contains an odd number of particles 
by one free (nonclustered) ion. 
With increasing Bjerrum length, the fraction of free ions decreases. 

We now assume that only free particles 
contribute to screening and split the number density $\rho_{\textrm{s}}$ of 
all ions into free and bound parts, 
$\rho_{\textrm{s}}=\rho_{\textrm{f}}+\rho_{\textrm{b}}$. 
Assuming only the free ions cause Debye screening, the expected decay length is 
simply the Debye length for the reduced density $\rho_{\textrm{f}}$, 
$\lambda^{\textrm{MD,f}}=1/\sqrt{8\pi\lambda_{\textrm{B}}\rho_{\textrm{f}}}$. 
In \cref{fig:cluster-decay}(a), the decay lengths resulting from this 
cluster analysis on the same simulation data that led to the results 
of \cref{fig:decay} are displayed for different 
connectivity shell diameters $d=1.5\sigma,2\sigma,3\sigma$ alongside the 
directly measured decay length $\lambda_2$ from \cref{fig:decay}.
Our cluster analysis predicts anomalous underscreening 
very similar to the direct extraction of decay lengths 
from simulation in \cref{fig:decay}. 
It even predicts anomalous underscreening for data points $(c,\lambda_{\mathrm{B}})$ 
where we could not use the direct fitting method due to insufficient numerical resolution. 
The cluster analysis also shows anomalous underscreening in 
\cref{fig:cluster-decay}(b) for the same parameter pairs 
$(c,\lambda_{\mathrm{B}})$ as used in the experiments of \cite{smith_jpcl7_2016}. With 
an adequate choice of connectivity diameter, this prediction 
even matches the experimentally measured decay lengths. 
However, while the predicted decay length is rather insensitive to the 
choice of connectivity diameter $d$ at low concentrations, which is a necessary condition for a 
meaningful trend as $d$ itself has no physical footing, 
at higher concentrations the choice of the connectivity diameter 
matters, rendering the method inapplicable. A better definition of free and 
bound ions might be facilitated by machine-learned local 
structures \cite{jones_jcp154_2021,jones_arxiv_2022}. 
Nevertheless, our cluster analysis 
supports the hypothesis that anomalous underscreening is also 
present at high concentrations in our MD simulations, but its signal is too 
small to be distinguished from noise \cite{Note1}.

To supplement our explanation of anomalous underscreening, we present a 
minimal theory that allows ion pairing, similar to previous 
approaches \cite{zwanikken_jpcm21_2009,adar_jcp146_2017}. 
We acknowledge that the general mechanism is presumably ``not a question of pair formation, 
but a more general transient association of ions involving several ions of 
opposite charge'' \cite{kjellander_jcp148_2018}. 
Our approach is based on the grand canonical description of an electrolyte of 
positive and negative point charges in a volume $V$, where particles either are free or bound in 
neutral pairs, 
$\beta\Omega^{\textrm{pair}}/V = 
2\rho_{\textrm{f}}(\log(\rho_{\textrm{f}}\Lambda_{\textrm{f}}^3)-1)
+ \rho_{\textrm{p}}(\log(\rho_{\textrm{p}}\Lambda_{\textrm{p}}^3)-1) 
+ F^{\textrm{es}} - \beta\mu_{\textrm{f}}\rho_{\textrm{f}} - \beta\mu_{\textrm{p}}\rho_{\textrm{p}}$. 
We eliminate the thermal wavelengths by identifying 
$\Lambda_{\textrm{s}}=\Lambda_{\textrm{f}}=\Lambda_{\textrm{p}}\sqrt{2}$ and 
comparing with a system of solely pointlike ions. 
Using $3/2 k_{\textrm{B}}T$ and the electrostatic bulk energy density 
$F^{\textrm{es}}=-\lambda_{\textrm{D}}^{-3}/(12\pi)$ \cite{debye_pz24_1923} 
for the inner energy per volume in units of $k_{\textrm{B}}T$, we obtain our final result 
\begin{align}
\beta \frac{\Omega^{\textrm{pair}}}{V} =& 
2\rho_{\textrm{f}}\left(\log(\rho_{\textrm{f}}/\rho_{\textrm{s}})-1\right)
+ \rho_{\textrm{p}}\left(\log(\rho_{\textrm{p}}/(\sqrt{2}^3\rho_{\textrm{s}})-1\right) \nonumber \\
&- \left( 1-\frac{3}{2}\frac{\rho_{\textrm{f}}}{\rho_{\textrm{s}}} \right) 
\frac{\sqrt{8\pi\lambda_{\textrm{B}}\rho_{\textrm{s}}}^3}{12\pi} 
+ \tfrac{3}{2}\rho_{\textrm{p}} , 
\label{eq:omega-pair-3}
\end{align}
as derived in more detail in the Supplemental Material \cite{Note1}. 
Setting $\rho_{\textrm{f}}=\alpha\rho_{\textrm{s}}$ and $\rho_{\textrm{p}}=(1-\alpha)\rho_{\textrm{s}}$ 
in \cref{eq:omega-pair-3}, we obtain $\Omega^{\textrm{pair}}(\alpha)$ with $\alpha\in[0,1]$ that can 
be minimized with respect to the fraction $\alpha$ of free ions while $\rho_{\textrm{s}}$ is kept fixed. 

As previously, in the cluster analysis of our simulation results, we assume 
that only free ions contribute to the screening of charges. 
Accordingly, we use the predicted density $\rho_{\textrm{f}}^{\textrm{theory}}$ of free ions 
to obtain the decay length 
$\lambda^{\textrm{theory}}=1/\sqrt{8\pi\lambda_{\textrm{B}}\rho_{\textrm{f}}^{\textrm{theory}}}$ 
as a function of the total ion concentration $\rho_{\textrm{s}}$ and the Bjerrum 
length $\lambda_{\textrm{B}}$. 
In \cref{fig:cluster-decay}, we sketch the predictions of this theory of ion pairing 
in comparison to our results from MD simulations and experimental data. 
Clearly, our minimal theory predicts an even stronger increase of the decay 
length than is found in simulations or experiments. 
While this increase starts at lower $\sigma/\lambda_{\textrm{D}}$ than expected 
[see \cref{fig:cluster-decay}(a)], the theory confirms the shift to 
larger $\sigma/\lambda_{\textrm{D}}$ with increasing ion concentration. 
In \cref{fig:cluster-decay}(b), the theory reproduces the strong increase 
of the experimentally reported decay lengths and its position in the plot 
remarkably well.

In summary, we show that anomalous underscreening, which previously has only 
been reported experimentally, can also be found in the RPM using MD 
simulations. Our results demonstrate that the decay length is, in general, not 
a unique function of the parameter $\sigma/\lambda_{\textrm{D}}$ as suggested 
by experiments \cite{lee_prl119_2017}, but the experiments probe only a 
unique line in the phase diagram of the RPM. 
On top of that, we illustrate that cluster formation induces a strong increase 
of the screening length, which provides an explanation for anomalous 
underscreening. 
We support this explanation, on the one hand, by analyzing clusters in our MD 
simulations and, on the other hand, by applying a minimal cluster 
theory of ion pairing which allows ions to form neutral pairs. 

Finally, the question remains why some experiments could find anomalous 
underscreening and others could not. 
As a possible answer, it has been proposed that the atomic force microscope 
has by construction a much lower sensitivity than the SFA 
\cite{gebbie_pnas110_2013,baimpos_langmuir30_2014}. 
Accordingly, the signal of anomalous underscreening might be too small for 
some of the experiments, similar to the sensitivity of our MD simulations 
\cite{Note1}.

\section*{Acknowledgements}

We thank Patrick Warren, Fabian Glatzel, and Anja Kuhnhold for fruitful discussions. 
AH and MB acknowledge funding from the German Research Foundation (DFG) through Project No. 406121234. 
FC acknowledges funding from the German Research Foundation (DFG) through Project No. 457534544. 
We acknowledge support by the state of Baden-W\"urttemberg through bwHPC and the German Research 
Foundation (DFG) through Grant No. INST 39/963-1 FUGG (bwForCluster NEMO).

%


\clearpage
\pagebreak

\makeatletter
\def\mysequence#1{\expandafter\@mysequence\csname c@#1\endcsname}
\def\@mysequence#1{%
  \ifcase#1\or 1\or 1\or 1\or 1\or 1\or 2\or 3\or 4\or 5\or 6\or 7\or 8\or 9\or 10\or 11\or 12\or 13\or 14\or 15\or 16\or 17\or 18\or 19\or 20\or 21\or 22\or 23\or 24\or 25\or 26\or 27\or 28\else\@ctrerr\fi}
\makeatother

\setcounter{equation}{0}
\setcounter{table}{0}
\setcounter{page}{1}
\makeatletter
\renewcommand{\theequation}{S\arabic{equation}}
\renewcommand{\thefigure}{S\mysequence{figure}}
\renewcommand{\bibnumfmt}[1]{[S#1]}
\renewcommand{\citenumfont}[1]{S#1}
\renewcommand{\thetable}{S\arabic{table}}

\title{Supplemental Material \\ for: Anomalous Underscreening in the Restricted Primitive Model}




\date[]{published as \href{http://link.aps.org/supplemental/10.1103/PhysRevLett.130.108202}{Supplemental Material} for: Physical Review Letters \textbf{130}, 108202 (2023), DOI: \href{https://doi.org/10.1103/PhysRevLett.130.108202}{10.1103/PhysRevLett.130.108202} }

\begin{abstract}
In this supplemental material for our article ``Anomalous Underscreening in the Restricted Primitive Model'' 
we present details on the (I.) minimal theory of ion pairing we used, 
on the (II.) Molecular Dynamics simulations we performed, 
on the (III.) cluster analysis of our simulation results, 
on the (IV.) theoretical foundations of decay lengths, and on the 
extraction of decay lengths from (V.) fitting of charge-correlation functions. 
We further show (VI.) additional results from density functional theory 
and list in \cref{Stab:exp-decay} the parameters 
and the corresponding measured decay lengths 
as reported for experimental measurements 
in the supporting information of \cite{Ssmith_jpcl7_2016}. 
\end{abstract}

\maketitle

\section{Minimal theory of ion pairing}

We consider a system of ideal positive and negative charges in a volume $V$ 
at temperature $T$. Each species has a number density $\rho_{\textrm{s}}$. 
With three degrees of freedom per particle, the internal energy $U$ of this system is 
\begin{align}
\beta\frac{U}{V} = \tfrac{3}{2}2\rho_{\textrm{s}} + F^{\textrm{es}} , 
\label{Seq:inner-energy}
\end{align}
with the bulk electrostatic energy 
$F^{\textrm{es}}=-\sqrt{8\pi\lambda_{\textrm{B}}\rho_{\textrm{s}}}^3/(12\pi)$ 
as derived by Debye and H\"uckel \cite{Sdebye_pz24_1923} and the inverse temperature 
$\beta=1/k_{\textrm{B}}T$ setting the thermal energy. 
As in the main article, $k_{\textrm{B}}$ is Boltzmann's constant and 
$\lambda_{\textrm{B}}$ the Bjerrum length. 
The chemical potential follows from \cref{Seq:inner-energy} with 
\begin{align}
\beta\mu_{\textrm{s}}^{\textrm{TD}} 
= \left(\frac{\partial\beta\tfrac{U}{V}}{\partial\rho_{\textrm{s}}}\right)_{S,V}
= 3 + \tfrac{3}{2} F^{\textrm{es}} \frac{1}{\rho_{\textrm{s}}} , 
\label{Seq:chempot-td}
\end{align}
where the entropy $S$ is a thermodynamic state variable. 
From another perspective, the grand potential $\Omega$ of the system reads 
\begin{align}
\beta \frac{\Omega}{V} = 2\rho_{\textrm{s}}\left(\log(\rho_{\textrm{s}}\Lambda_{\textrm{s}}^3)-1\right)
+ F^{\textrm{es}} - \beta\mu_{\textrm{s}}\rho_{\textrm{s}}  
\end{align}
with the thermal de Broglie wavelength $\Lambda_{\textrm{s}}$. 
As in classical density functional theory (DFT) \cite{Shansen_book_2013}, the variational principle 
\begin{align}
\frac{\delta \beta\tfrac{\Omega}{V}}{\delta\rho_{\textrm{s}}} 
= 2\log(\rho_{\textrm{s}}\Lambda_{\textrm{s}}^3) 
  + \tfrac{3}{2} F^{\textrm{es}} \frac{1}{\rho_{\textrm{s}}} - \beta\mu_{\textrm{s}}
\stackrel{!}{=} 0 
\label{Seq:chempot-dft}
\end{align}
holds for the true physical system in equilibrium. 

As we aim for a description of free and paired ions, we now consider three species, 
namely positive and negative free ions with number density $\rho_{\textrm{f}}$ and 
neutral pairs of ions with number density $\rho_{\textrm{p}}$. The total 
number of ions is conserved by $\rho_{\textrm{s}}=\rho_{\textrm{f}}+\rho_{\textrm{p}}$. 
The grand potential of this system is given by
\begin{align}
\beta \frac{\Omega^{\textrm{pair}}}{V} =& 
2\rho_{\textrm{f}}\left(\log(\rho_{\textrm{f}}\Lambda_{\textrm{f}}^3)-1\right)
+ \rho_{\textrm{p}}\left(\log(\rho_{\textrm{p}}\Lambda_{\textrm{p}}^3)-1\right) \nonumber \\
 &+ F^{\textrm{es}} - \beta\mu_{\textrm{f}}\rho_{\textrm{f}} - \beta\mu_{\textrm{p}}\rho_{\textrm{p}} . 
\label{Seq:omega-pair-1}
\end{align}
Free ions have the same thermal wavelength as the salt but a pair has twice the mass 
of a free ion leading to $\Lambda_{\textrm{s}}=\Lambda_{\textrm{f}}=\Lambda_{\textrm{p}}\sqrt{2}$. 
Furthermore, the internal energy of the system does not change with changing the number of pairs, thus, 
we identify $\mu_{\textrm{f}}=\mu_{\textrm{s}}$ and $\mu_{\textrm{p}}=0$. 
Combining \cref{Seq:omega-pair-1,Seq:chempot-dft} we can eliminate the thermal wavelengths. 
In a last step we identify $\mu_{\textrm{s}}=\mu_{\textrm{s}}^{\textrm{TD}}$ and 
insert \cref{Seq:chempot-td} yielding our final result 
\begin{align}
\beta \frac{\Omega^{\textrm{pair}}}{V} =& 
2\rho_{\textrm{f}}\left(\log(\rho_{\textrm{f}}/\rho_{\textrm{s}})-1\right)
+ \rho_{\textrm{p}}\left(\log(\rho_{\textrm{p}}/(\sqrt{2}^3\rho_{\textrm{s}})-1\right) \nonumber \\
&+ \left( 1-\frac{3}{2}\frac{\rho_{\textrm{f}}}{\rho_{\textrm{s}}} \right)F^{\textrm{es}}
+ \tfrac{3}{2}\rho_{\textrm{p}} . 
\label{Seq:omega-pair-3}
\end{align}
Setting $\rho_{\textrm{f}}=\alpha\rho_{\textrm{s}}$ and $\rho_{\textrm{p}}=(1-\alpha)\rho_{\textrm{s}}$ 
in \cref{Seq:omega-pair-3}, we obtain $\Omega^{\textrm{pair}}(\alpha)$ with $\alpha\in[0,1]$ that can 
be minimized with respect to the fraction $\alpha$ of free ions while $\rho_{\textrm{s}}$ is kept fixed.

\section{Simulation details}

We consider the RPM of charged hard spheres with ion diameter $\sigma$ and valencies $Z_{\pm}=\pm 1$. 
The bulk system at temperature $T=293.41$ K 
has a thermal energy $\beta^{-1}=k_{\textrm{B}}T$ with Boltzmann's constant $k_{\textrm{B}}$. 
As in previous work \cite{Scats_jcp154_2021}, we perform extensive Molecular Dynamics (MD) 
simulations using the ESPResSo package \cite{Sweik_epjst227_2019}. 
Namely, we performed MD simulations of the RPM with different ion concentrations $c$ 
and Bjerrum lengths $\lambda_{\textrm{B}}$, while in most cases the ion diameter was 
set to $\sigma=0.3$ nm. 

In our MD simulation we measure energy in $k_{\textrm{B}}$T and distances in nm
which in combination with the mass $3\cdot 10^{-23}$ g of each of all the particles in the system 
defines the characteristic time scale of $2.699$ ps. 
We model the hard steric repulsion between ions by a Weeks-Chandler-Anderson potential 
\cite{Sandersen_pra4_1971,Sweeks_jcp54_1971} 
\begin{align}
u_{\textrm{WCA}}(r) = \left\{ \begin{matrix}4\gamma\left(\left(\frac{\sigma_{\textrm{LJ}}}{r}\right)^{12}
  -\left(\frac{\sigma_{\textrm{LJ}}}{r}\right)^6 + \frac{1}{4}\right) & r<\sigma \\
  0 & r\geq\sigma \end{matrix} \right.
\end{align}
with $\gamma=5 \cdot 10^3$ $k_{\textrm{B}}T$ and $\sigma_{\textrm{LJ}}=2^{-1/6} \sigma$ such that 
the potential is purely repulsive and its derivative is continuous at $\sigma$. To compute the electrostatic forces between ions, we use the 
P3M method, a sophisticated Ewald summation technique implemented in ESPResSo 
\cite{Shockney_book_1988,Sweik_epjst227_2019}. The system is coupled to a heat bath via a Langevin thermostat.
We simulate a bulk system by calculating ion trajectories in an almost cubic simulation box of 
volume $L_{\textrm{x}}\times L_{\textrm{y}}\times L_{\textrm{z}}$ with periodic boundary conditions. 
After setting $L_{\textrm{x}}=L_{\textrm{y}}=L_{\textrm{z}}$ and the concentration $c$ of 
positive and negative particles, respectively, we slightly have to adjust $L_{\textrm{x}}$ and 
$L_{\textrm{y}}$ such that an integer number of particles fits into the simulation box at the 
chosen $L_{\textrm{z}}$ and $c$. 
Note that instead of the concentration, equivalently often the number density $\rho_{\textrm{s}}$ of each ion species 
is used. 

Each simulation is initiated by a random configuration that is relaxed by slowly increasing the 
repulsive inter-particle potential $u_{\textrm{WCA}}$. After switching on the additional electrostatic 
particle interactions, the system is evolved until the 
energy of the system fluctuates on a stable level. 
Then we start sampling radial pair-distribution functions $g_{\mu\nu}(r)$ between particles 
of species $\mu$ and $\nu$, respectively using a histogram
featuring 400 equidistant bins covering the separation interval $r\in[0,5]$ nm in order to achieve a high resolution. 
Between two configurations analyzed in this way we perform 1000 integration steps 
to decorrelate the respective system snapshots.   
To achieve better statistics we perform several independent simulation runs of the same 
system in parallel. 
In \cref{Stab:sim-details} we list concentration $c$, Bjerrum length $\lambda_{\textrm{B}}$, ion diameter $\sigma$, 
system size $L_{\textrm{z}}$, and the total number of configurations $N_{\textrm{g}}$ used 
for sampling a radial pair-distribution function for all simulations analyzed. 
A typical system size is $L_{\textrm{z}}=60$ nm and the number of 
configurations that the radial distribution functions are averaged over ranges from $10^3$ up to $10^6$, 
depending on system size and the concentration. 
The number of particles that we used for a system size of $L=60$ nm is summarized 
in \cref{Stab:particle-numbers} for concentrations $c=0.05$ mol/L, $0.1$ mol/L, $0.2$ mol/L, $0.5$ mol/L 
and can be extracted for all simulated systems from the parameters listed in \cref{Stab:sim-details}. 

\begin{table}[ht]
\caption{\label{Stab:particle-numbers} Size $L_x\times L_y\times L_z$ of the simulation 
box and number $N_{\pm}$ of positive and negative ions, respectively, at a given concentration $c$. 
$N_+=N_-$ ensures overall charge neutrality. }
\centering
\begin{tabular}{|l|l|l|l|l|}
\hline
 $c$ (mol/L) & $L_x=L_y$ (nm) & $L_z$ (nm) & $N_{\pm}$ \\
\hline
0.05 & $\approx60.00111$ & 60 & 6504 \\
0.1  & $\approx60.00111$ & 60 & 13008 \\
0.2  & $\approx60.00111$ & 60 & 26016 \\
0.5  & $\approx60.00018$ & 60 & 65038 \\
\hline
\end{tabular}
\end{table}

\section{Cluster analysis of the simulation results}

\begin{figure}[t]
\centering
\includegraphics[width=\linewidth]{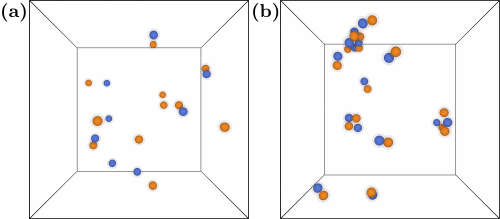}
\caption{Snippets from particle configurations obtained from MD simulations 
at $c=0.1$ mol/L and (a) $\lambda_{\textrm{B}}=0.7$ nm and (b) $\lambda_{\textrm{B}}=5$ nm. 
The particle diameter is $\sigma=0.3$ nm and the connectivity diameter is $d=1.5\sigma$. 
Each snippet has the size $6\times 6\times 5$ nm$^3$, where the depth is $5$ nm. 
}
\label{Sfig:snapshot}
\end{figure}

For increasing Bjerrum length we observed a clustering of ions in our MD simulation results. 
An example is illustrated in \cref{Sfig:snapshot} for a system at $c=0.1$ mol/L. 
The figure shows two identically sized snippets from systems simulated at 
different Bjerrum lengths (a) $\lambda_{\textrm{B}}=0.7$ nm and 
(b) $\lambda_{\textrm{B}}=5$ nm. At higher Bjerrum length we observe a multitude of clusters 
with even number of ions. Typically, the charge of all ions in such an even cluster sums up to zero, 
as we will demonstrate in the following.

\begin{figure}[t]
\centering
\includegraphics[width=\linewidth]{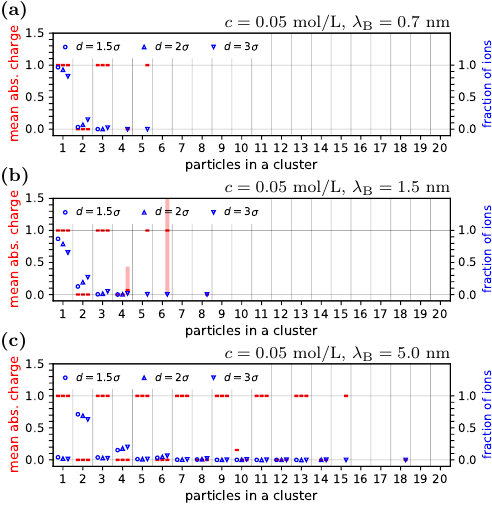}
\caption{Mean absolute charge (in elementary charges $e$) of clusters that contain 
$N_{\textrm{c}}$ particles (red bars, left axis) and fraction of ions in all cluster of the 
respective size with respect to the number of ions in the system (blue symbols, right axis). 
The data depends on the connectivity length $d$, 
thus, on the definition of a cluster. 
For each cluster size $N_{\textrm{c}}$ with a respective number of particles in a cluster 
three values for $d=1.5\sigma$, $2\sigma$, $3\sigma$ are shown. 
Averages were taken for all clusters found in $10$ spatial configuration snapshots of a 
simulation with concentration $c=0.05$ mol/L and Bjerrum length 
(a) $\lambda_{\textrm{B}}=0.7$ nm, (b) $1.5$ nm, and (c) $5.0$ nm. 
Bars indicate the standard deviation from the absolute mean charge. 
If no clusters of a certain size were found, no data points are shown. }
\label{Sfig:cluster-analysis3-c0.05}
\end{figure}

\begin{figure}[t]
\centering
\includegraphics[width=\linewidth]{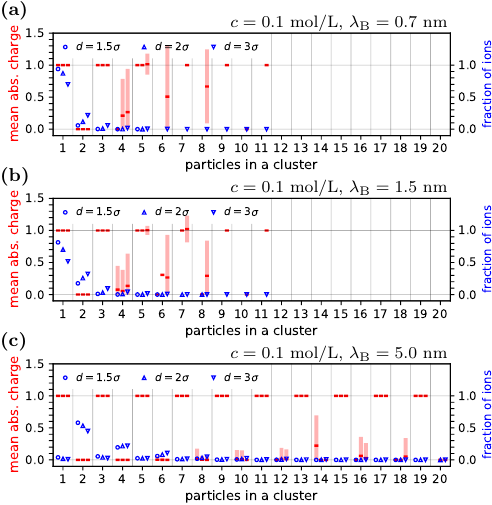}
\caption{Mean absolute charge of clusters (in elementary charges $e$) and 
fraction of all ions in a cluster 
as described in the caption of \cref{Sfig:cluster-analysis3-c0.05} 
for systems of concentration $c=0.1$ mol/L and Bjerrum length 
(a) $\lambda_{\textrm{B}}=0.7$ nm, (b) $1.5$ nm, and (c) $5.0$ nm. }
\label{Sfig:cluster-analysis3-c0.1}
\end{figure}

\begin{figure}[t]
\centering
\includegraphics[width=\linewidth]{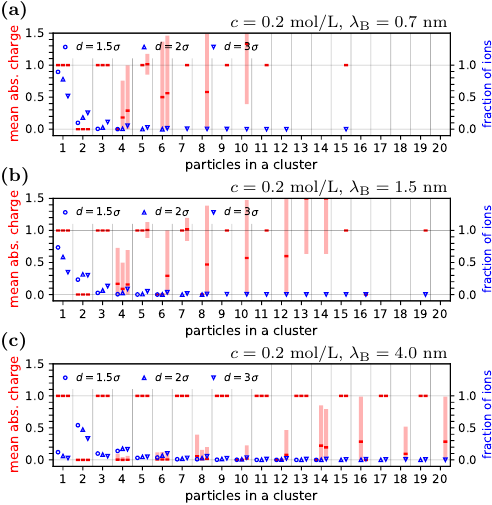}
\caption{Mean absolute charge of clusters (in elementary charges $e$) and fraction of all ions in a cluster 
as described in the caption of \cref{Sfig:cluster-analysis3-c0.05}
for systems of concentration $c=0.2$ mol/L and Bjerrum length
(a) $\lambda_{\textrm{B}}=0.7$ nm, (b) $1.5$ nm, and (c) $4.0$ nm; in (c) we show data for 
$\lambda_{\textrm{B}}=4$ nm instead of $5$ nm. 
}
\label{Sfig:cluster-analysis3-c0.2}
\end{figure}

\begin{figure}[t]
\centering
\includegraphics[width=\linewidth]{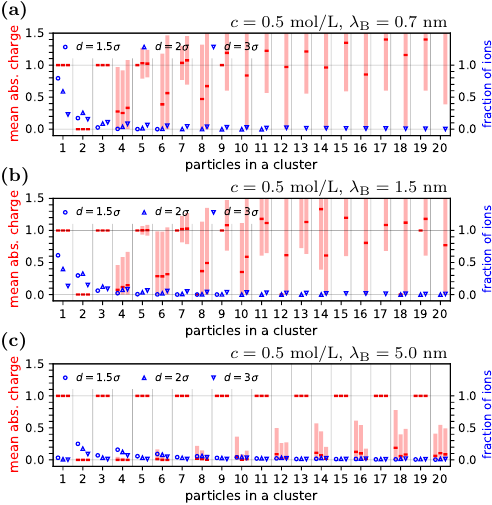}
\caption{Mean absolute charge of clusters (in elementary charges $e$) and fraction of all ions in a cluster 
as described in the caption of \cref{Sfig:cluster-analysis3-c0.05}
for systems of concentration $c=0.5$ mol/L and Bjerrum length
(a) $\lambda_{\textrm{B}}=0.7$ nm, (b) $1.5$ nm, and (c) $5.0$ nm. }
\label{Sfig:cluster-analysis3-c0.5}
\end{figure}

For a more detailed analysis we performed a cluster analysis on 
some of the particle configurations 
that we used to sample the pair-distribution functions. 
In this cluster analysis we identified particles in a configuration as 
connected or clustered if their distance to 
neighbouring particles was less than a connectivity distance $d>\sigma$. 
All particles that are connected with each other form a cluster. The size of a cluster is given 
by the number $N_{\textrm{c}}$ of particles forming it. 

An overview over the detected clusters and their average charge 
is presented in \cref{Sfig:cluster-analysis3-c0.05,Sfig:cluster-analysis3-c0.1,Sfig:cluster-analysis3-c0.2,Sfig:cluster-analysis3-c0.5}, 
where the figures show results for the concentrations 
$c=0.05$ mol/L, $0.1$ mol/L, $0.2$ mol/L, and $0.5$ mol/L, respectively. 
Each figure, again, contains three panels that show results for different Bjerrum lengths as indicated 
above each panel. 
In each panel, the number of particles in a cluster $N_{\textrm{c}}$ is given on the $x$-axis. 
For each cluster size $N_{\textrm{c}}$, again, data is shown for three different connectivity 
distances $d$, sorted from left to right and distinguished by three different symbols for the fraction 
of ions as indicated in each panel. 
If no symbol is shown, no clusters of the respective size $N_{\textrm{c}}$ have been found in the 
analyzed configurations for the given connectivity distance $d$. 

The \cref{Sfig:cluster-analysis3-c0.05,Sfig:cluster-analysis3-c0.1,Sfig:cluster-analysis3-c0.2,Sfig:cluster-analysis3-c0.5} show the average absolute charge on clusters of a given 
size $N_{\textrm{c}}$ by red bars with error bars that indicate the standards deviation (left $y$-axis). 
In addition, blue symbols indicate the fraction $n_{N_{\textrm{c}}}$ of ions in the system that are clustered in 
clusters of the respective size $N_{\textrm{c}}$. As the data shows, a small fraction 
$n_{N_{\textrm{c}}}$ comes with a huge standard deviation due to a small total number of available 
clusters of the respective size, for instance for a cluster of size $N_{\textrm{c}}=6$ in panel (b) 
of \cref{Sfig:cluster-analysis3-c0.05} ($c=0.05$ mol/L, $\lambda_{\textrm{B}}=1.5$ nm, $d=3\sigma$). 
The figures further demonstrate that the standard deviation of the average charge is larger for 
larger concentrations, which is expected for dense systems where the definition of a cluster is 
problematic. 

The analysis in \cref{Sfig:cluster-analysis3-c0.05,Sfig:cluster-analysis3-c0.1,Sfig:cluster-analysis3-c0.2,Sfig:cluster-analysis3-c0.5} shows that most clusters carry a net charge of $0$ or $\pm1$ elementary charges $e$. 
This holds in particular for low concentrations as well as for large Bjerrum lengths. 
Based on this finding, we simplify our cluster analysis and assume even clusters (clusters of even size) 
to be overall charge neutral and odd clusters (clusters of odd size) to carry on average only one 
positive or one negative net elementary charge $e$. 
Consequently, we also assume that even clusters do not contribute to global screening at all 
and odd clusters contribute to global screening as if they were free (not clustered) ions.

\begin{figure*}[t]
\centering
\includegraphics[width=\linewidth]{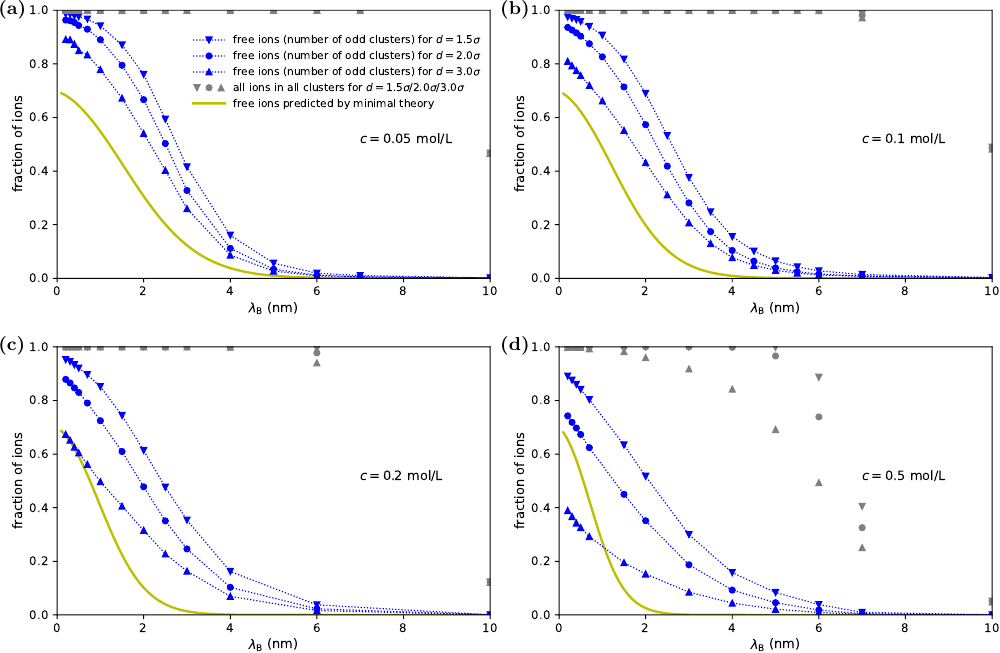}
\caption{Fraction of free ions in a system at concentration 
(a) $c=0.05$ mol/L, (b) $0.1$ mol/L, (c) $0.2$ mol/L, and (d) $0.5$ mol/L and Bjerrum length $\lambda_{\textrm{B}}$. 
Dotted lines guide the eyes. 
Gray symbols without dotted lines show the total fraction of particles in all clusters up to 
size $N_{\textrm{c}}=40$. 
The fraction of ions depends on the connectivity length $d$, thus, on the definition of a cluster. 
For each cluster size with $N_{\textrm{c}}$ particles in a cluster three values for $d=1.5\sigma$, 
$2\sigma$, $3\sigma$ are shown. 
Averages were taken for all clusters found in 10 spatial configuration snapshots of a 
simulation at respective parameters. 
The yellow solid line shows the prediction from the minimal theory of ion pairing. }
\label{Sfig:cluster-analysis}
\end{figure*}

In conclusion we define the number of odd clusters (not of the ions) as the number of free ions that 
still contribute to global screening. This definition includes all ions that are not clustered as well 
as the clusters that contain an odd number of ions such that the net charge of the cluster is that of 
a single ion. 
With increasing Bjerrum length the fraction of free ions decreases, which is 
shown in \cref{Sfig:cluster-analysis}. This result depends on the connectivity diameter 
$d$ that defines the distribution of clusters.

With increasing concentration, the cluster definition becomes problematic, 
because the choice of the connectivity distance determines the cluster size distribution. 
As the concentration is increased, the average number of particles in a cluster rises up to the point at 
which the entire system is spanned by one single cluster.

\section{Decay length theory}

At sufficiently large separations $r$, the charge-correlation function $h_{\textrm{cc}}(r)$ decays proportional 
to $\exp(-r/\lambda)/r$ \cite{Sattard_pra45_1992,Skjellander_jpl200_1992,Sevans_mp80_1993,
Sevans_jcp100_1994,SleoteDeCarvalho_mp83_1994} defining the screening length $\lambda$.
More precisely, in liquid-state theory the species dependent pair-correlation functions are related to the so-called
direct correlation functions via the Ornstein-Zernike 
equation \cite{Shansen_book_2013} which yields an algebraic matrix equation in Fourier space. 
In the RPM the eigenvalues of this matrix correspond to charge- and density-correlation function, respectively. 
Using contour integration, the inverse Fourier transform of the charge-correlation function can 
be expressed formally as a sum indexed by the set $Q_{\textrm{cc}}$ of all complex singularities of a 
rational function in the Fourier transforms of the direct correlation functions 
(see \cite{Scoupette_prl121_2018} for a brief description). 

Each simple pole $q_i\in Q_{\textrm{cc}}$ now adds a contribution $P_i(r)$ to the 
charge-correlation function $h_{\textrm{cc}}(r)$, which is of the form 
\begin{align}
P_i(r) = A_i \exp(-r/\lambda_i)\cos(\omega_i r+\tau_i)/r 
\label{Seq:pole-contribution}
\end{align}
with decay length $\lambda_i$, amplitude $A_i$,
and $\omega_i$ and $\tau_i$ describing potential oscillations.
In the limit of large separations, $r\to\infty$, all contributions become negligible compared to the 
term $P_i$ with the largest decay length $\lambda_i$. 
This asymptotically dominant decay length  of the charge-correlation function is the electrostatic screening length. 
In addition to the poles, a branch-point singularity can appear that leads to a slightly different 
decay contribution \cite{Sennis_jcp102_1995,Sulander_jcp114_2001}
\begin{align}
B_i(r) = A_i \exp(-r/\lambda_i)\cos(\omega_i r+\tau_i)/r^2 . 
\label{Seq:branch-contribution}
\end{align}

Notice, that analogous definitions can be made for 
density correlations $h_{\textrm{dd}}$ (also labeled particle-particle correlations) that lead to a similar decay 
length $\lambda_{\textrm{dd}}$ which may exceed the electrostatic screening length.
In the RPM, the longer of the two decay lengths dominates the decay 
of all pair-distribution functions $g_{\mu\nu}$ at long separations \cite{Scoupette_prl121_2018}. 
This gives rise to a potential crossover from 
charge-dominated to density-dominated decay \cite{Scoupette_prl121_2018}. 
Yet, charge and density correlations are related to the total correlation functions $h_{\mu\nu} := g_{\mu\nu} - 1$ via 
\begin{align}
	h_{\textrm{cc}} &= \frac{1}{2} \sum_{\mu}\sum_{\nu}Z_{\mu} Z_{\nu} h_{\mu\nu} , \\
	h_{\textrm{dd}} &= \frac{1}{2} \sum_{\mu}\sum_{\nu} h_{\mu\nu} . 
\end{align}
Charge inversion symmetry implies 
$h_{++}=h_{--}$ and $h_{+-}=h_{-+}$ such that the above equations reduce to 
$h_{\textrm{cc}}=h_{++}-h_{+-}$ and $h_{\textrm{dd}}=h_{++}+h_{+-}$ in the RPM. 
Thus, we can specifically compute $h_{\textrm{cc}} = g_{++}-g_{+-}$, the dominant decay of which always 
corresponds to the electrostatic screening length. Moreover, as long as the ion diameter is 
not significantly larger than the Bjerrum length, the asymptotic decay in the RPM is charge-driven.

\section{Fitting of charge-correlation functions}

\begin{figure*}[t]
\centering
\includegraphics[width=\linewidth]{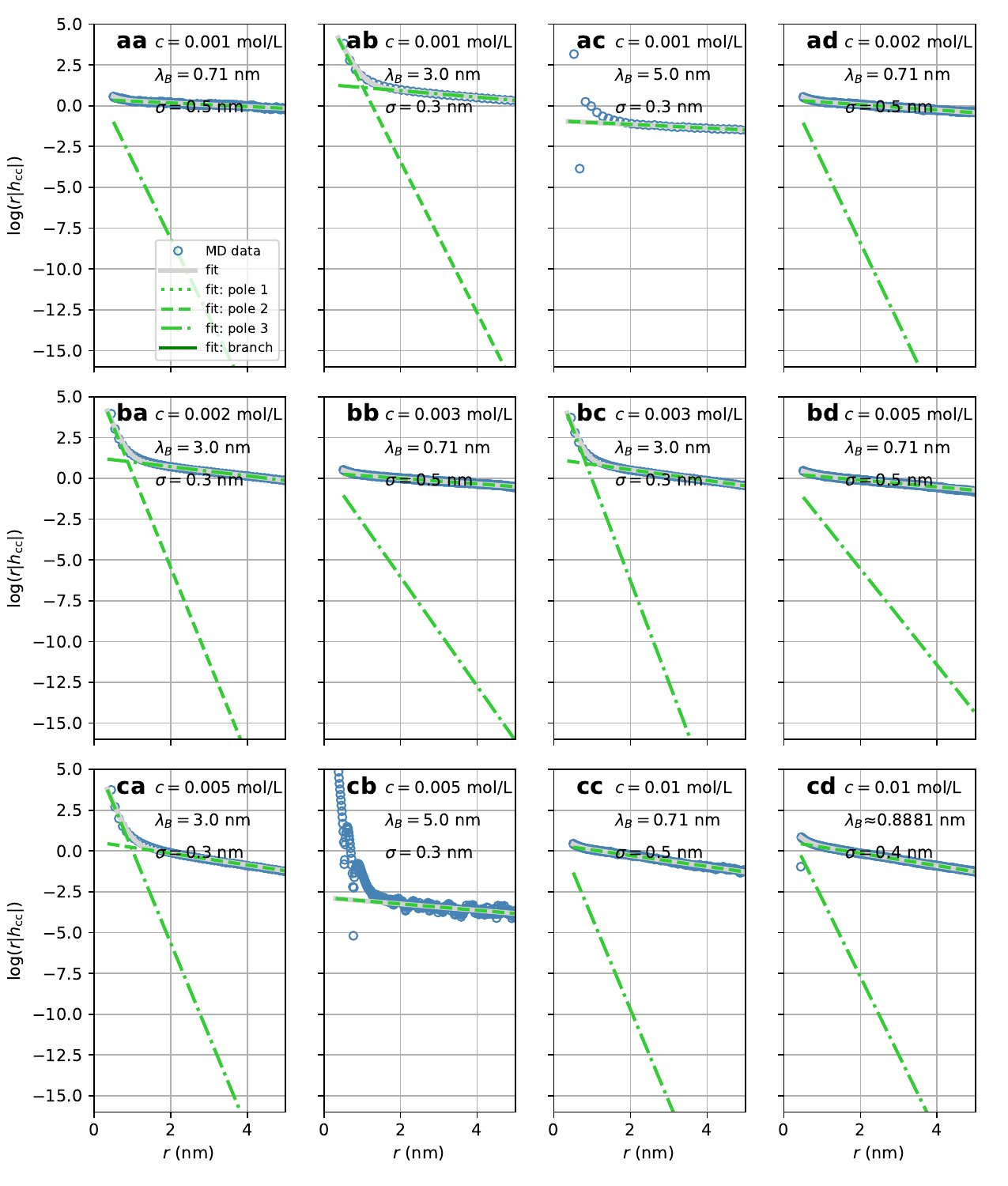}
\caption{Charge-correlation functions $h_{\textrm{cc}}(r)$ for parameter sets 
as indicated and listed in \cref{Stab:sim-details}. The pole and branch fits follow 
\cref{Seq:pole-contribution,Seq:branch-contribution} and their sum gives the 
thick, solid, gray fit function as explained in the text. }
\label{Sfig:corrfits01}
\end{figure*}

\begin{figure*}[t]
\centering
\includegraphics[width=\linewidth]{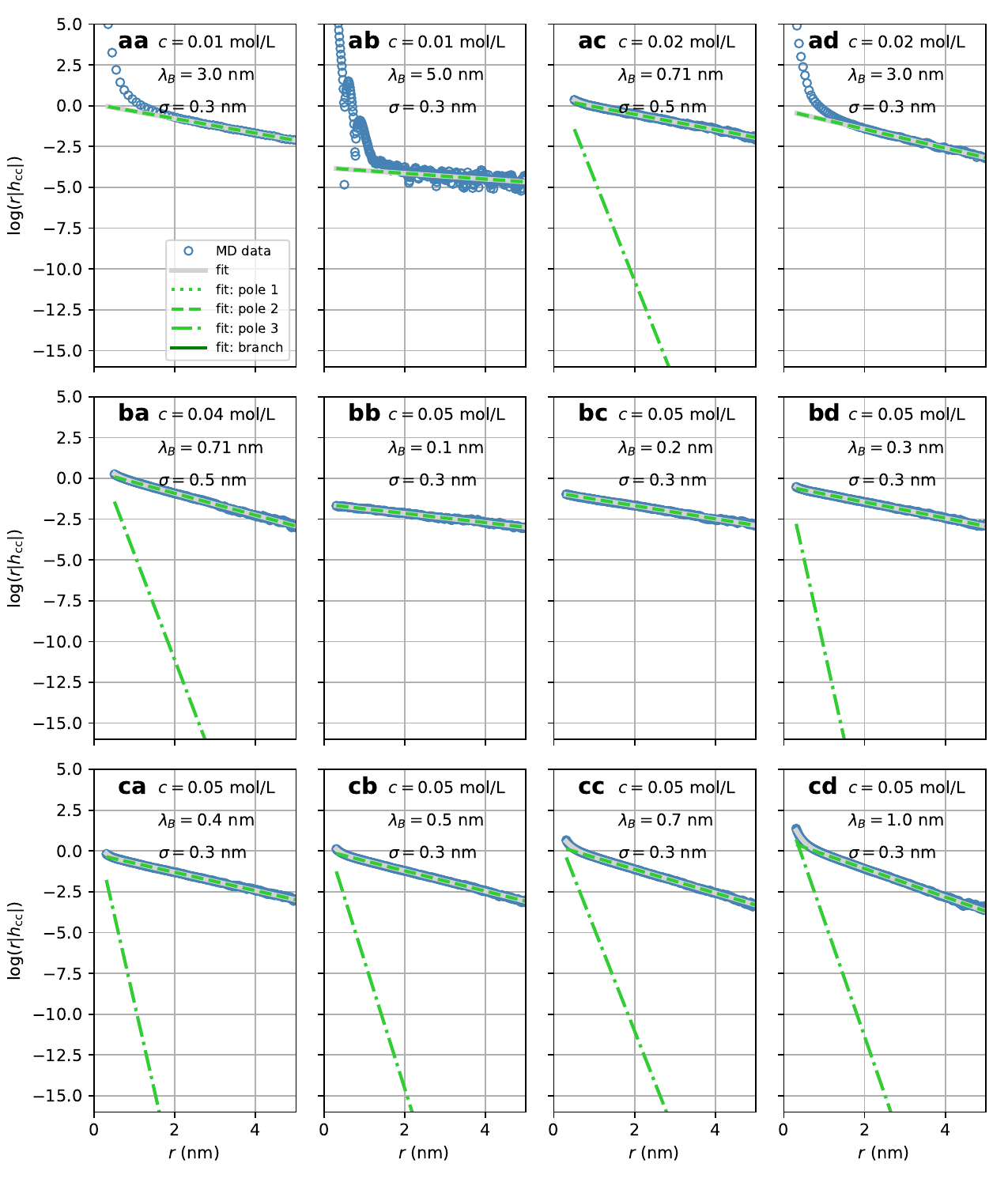}
\caption{Charge-correlation functions $h_{\textrm{cc}}(r)$ as explained in 
\cref{Sfig:corrfits01}, but for other parameter sets, as indicated and listed in \cref{Stab:sim-details}. }
\label{Sfig:corrfits02}
\end{figure*}

\begin{figure*}[t]
\centering
\includegraphics[width=\linewidth]{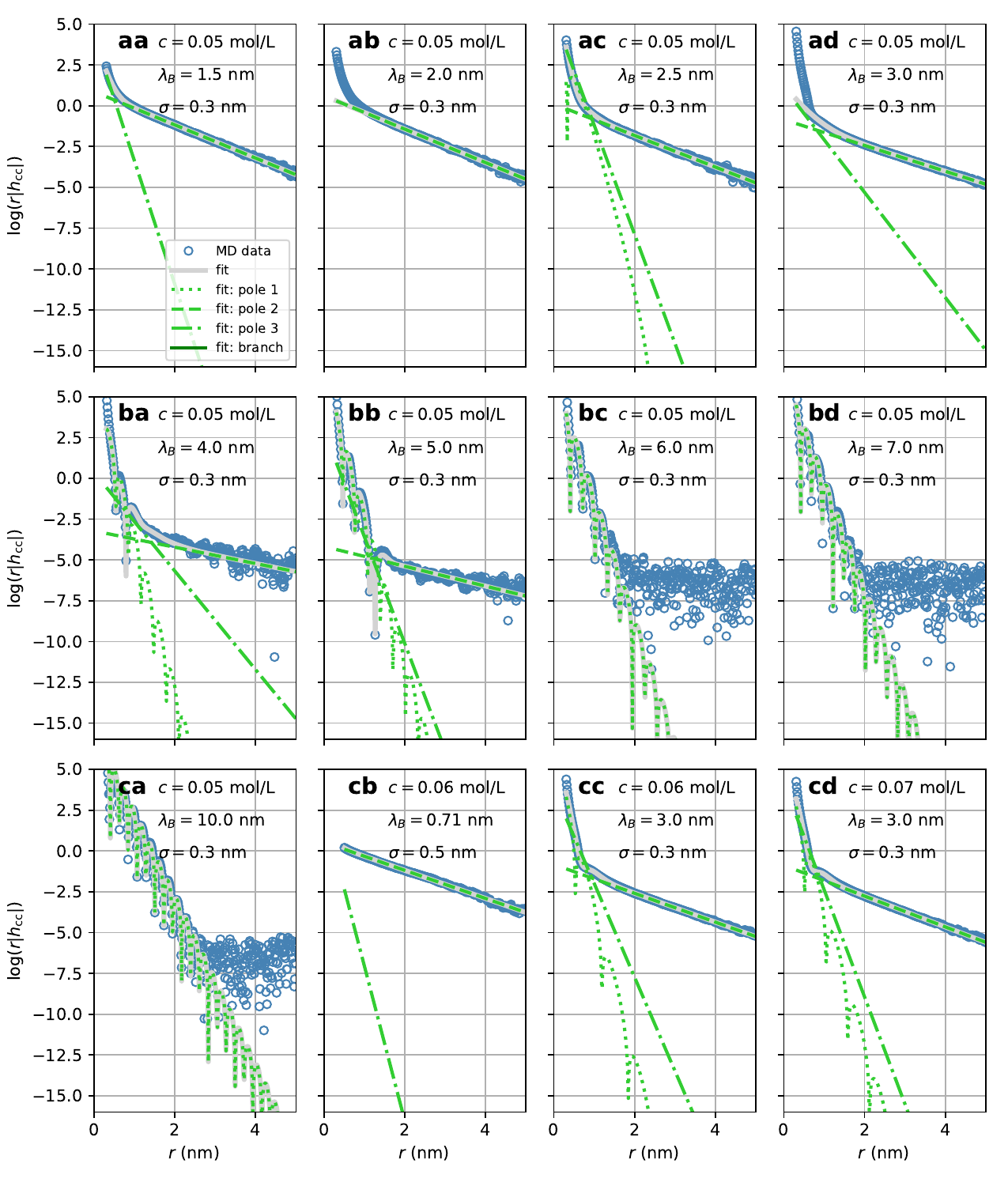}
\caption{Charge-correlation functions $h_{\textrm{cc}}(r)$ as explained in 
\cref{Sfig:corrfits01}, but for other parameter sets, as indicated and listed in \cref{Stab:sim-details}. }
\label{Sfig:corrfits03}
\end{figure*}

\begin{figure*}[t]
\centering
\includegraphics[width=\linewidth]{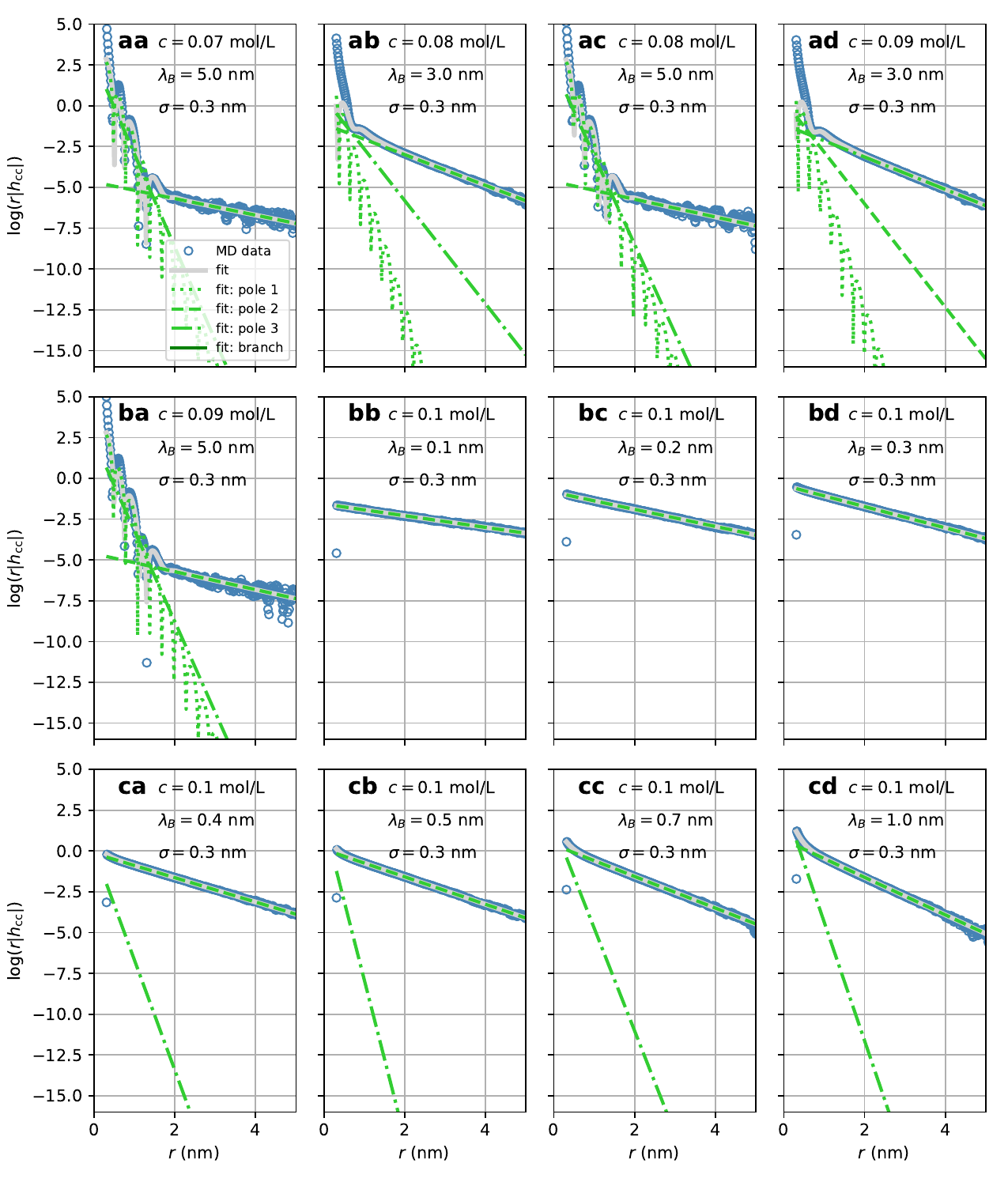}
\caption{Charge-correlation functions $h_{\textrm{cc}}(r)$ as explained in 
\cref{Sfig:corrfits01}, but for other parameter sets, as indicated and listed in \cref{Stab:sim-details}. }
\label{Sfig:corrfits04}
\end{figure*}

\begin{figure*}[t]
\centering
\includegraphics[width=\linewidth]{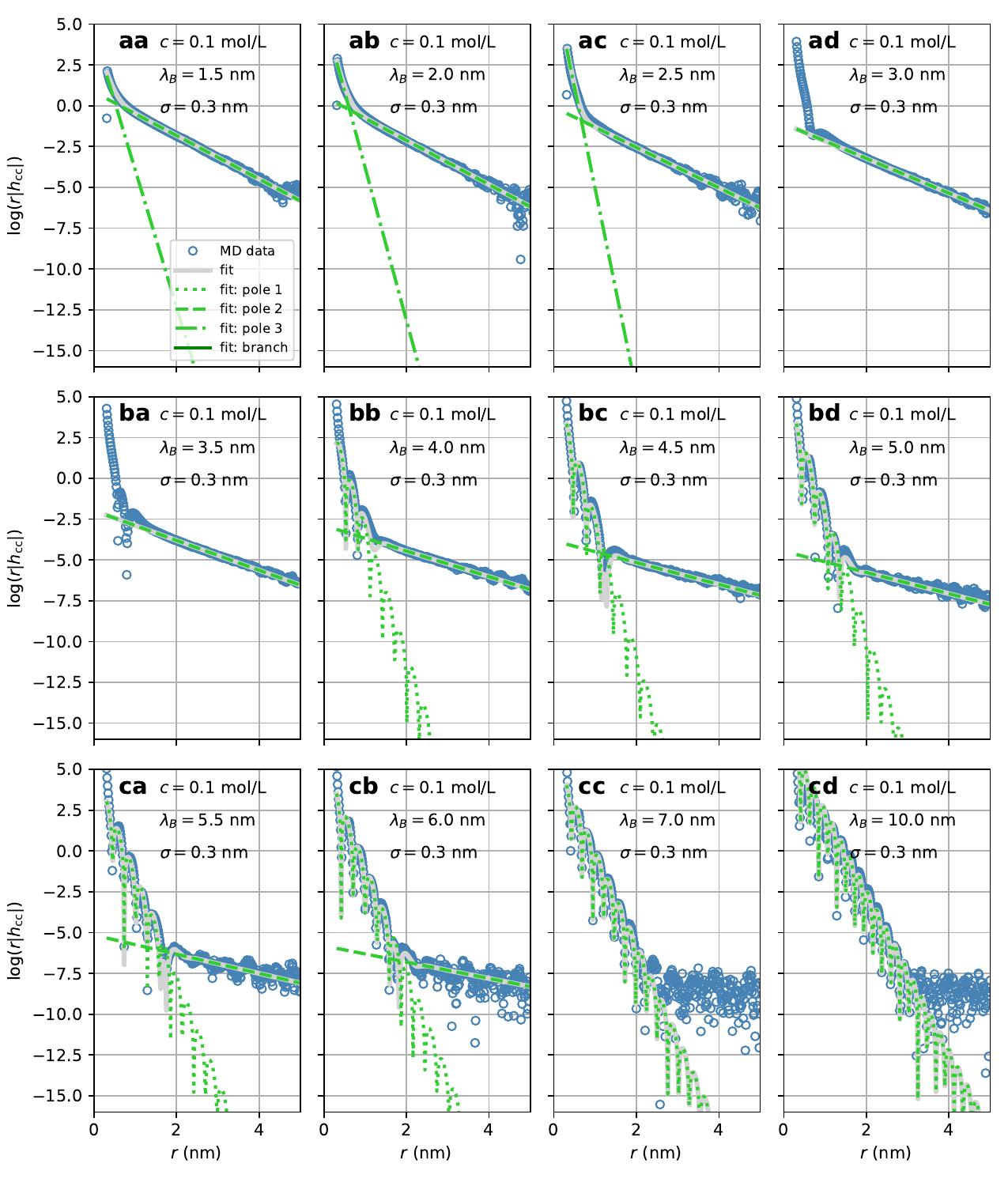}
\caption{Charge-correlation functions $h_{\textrm{cc}}(r)$ as explained in 
\cref{Sfig:corrfits01}, but for other parameter sets, as indicated and listed in \cref{Stab:sim-details}. }
\label{Sfig:corrfits05}
\end{figure*}

\begin{figure*}[t]
\centering
\includegraphics[width=\linewidth]{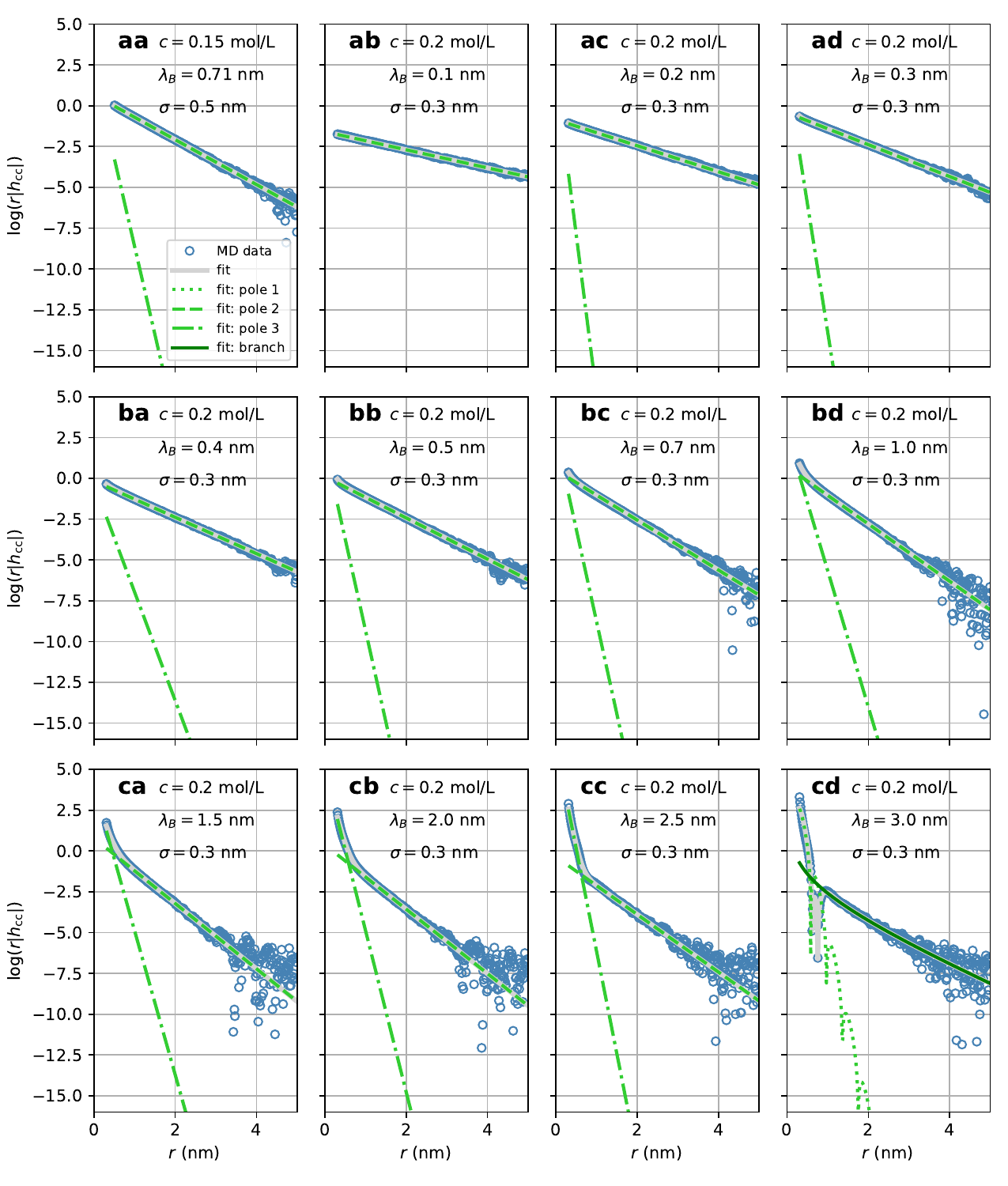}
\caption{Charge-correlation functions $h_{\textrm{cc}}(r)$ as explained in 
\cref{Sfig:corrfits01}, but for other parameter sets, as indicated and listed in \cref{Stab:sim-details}. }
\label{Sfig:corrfits06}
\end{figure*}

\begin{figure*}[t]
\centering
\includegraphics[width=\linewidth]{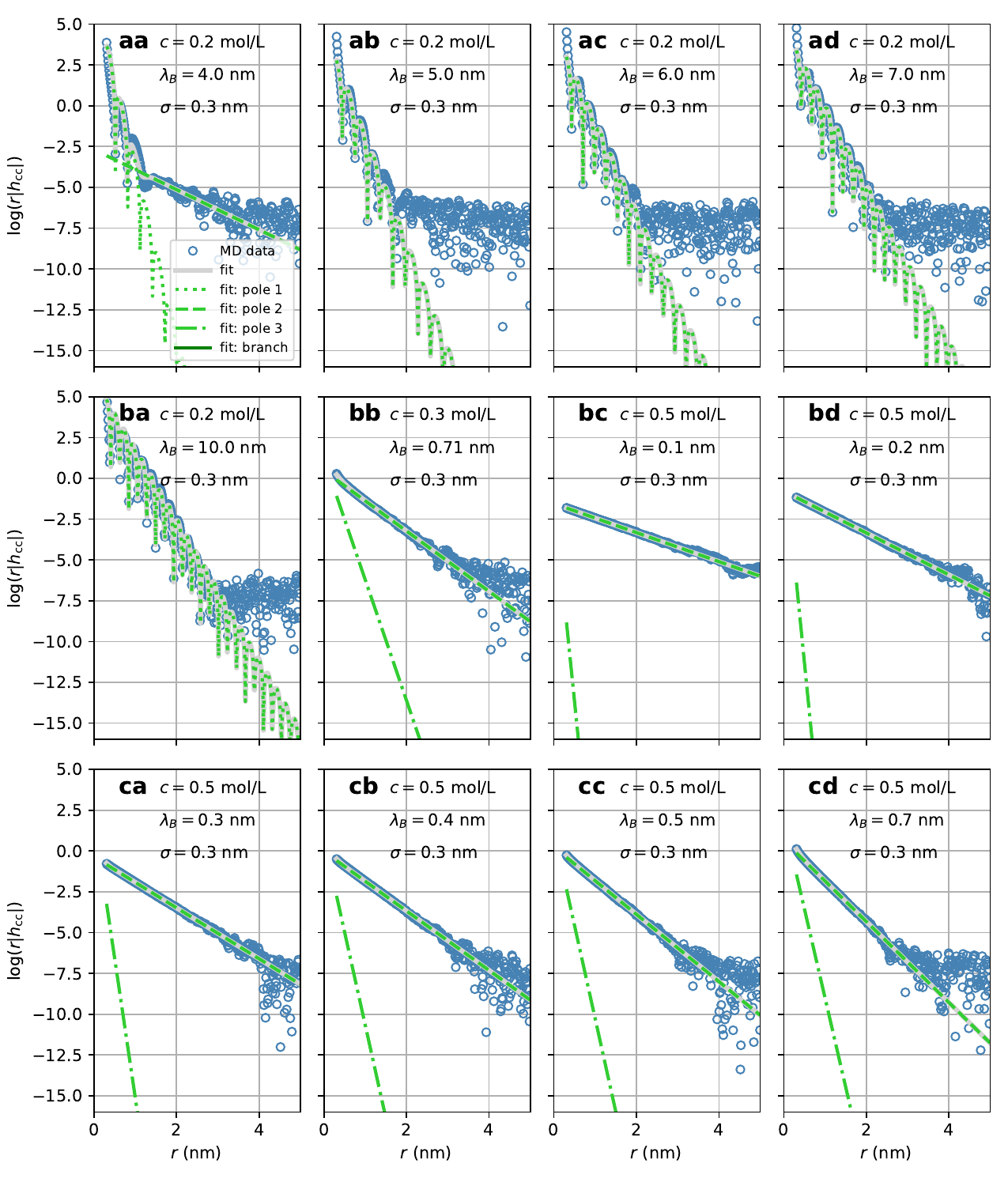}
\caption{Charge-correlation functions $h_{\textrm{cc}}(r)$ as explained in 
\cref{Sfig:corrfits01}, but for other parameter sets, as indicated and listed in \cref{Stab:sim-details}. }
\label{Sfig:corrfits07}
\end{figure*}

\begin{figure*}[t]
\centering
\includegraphics[width=\linewidth]{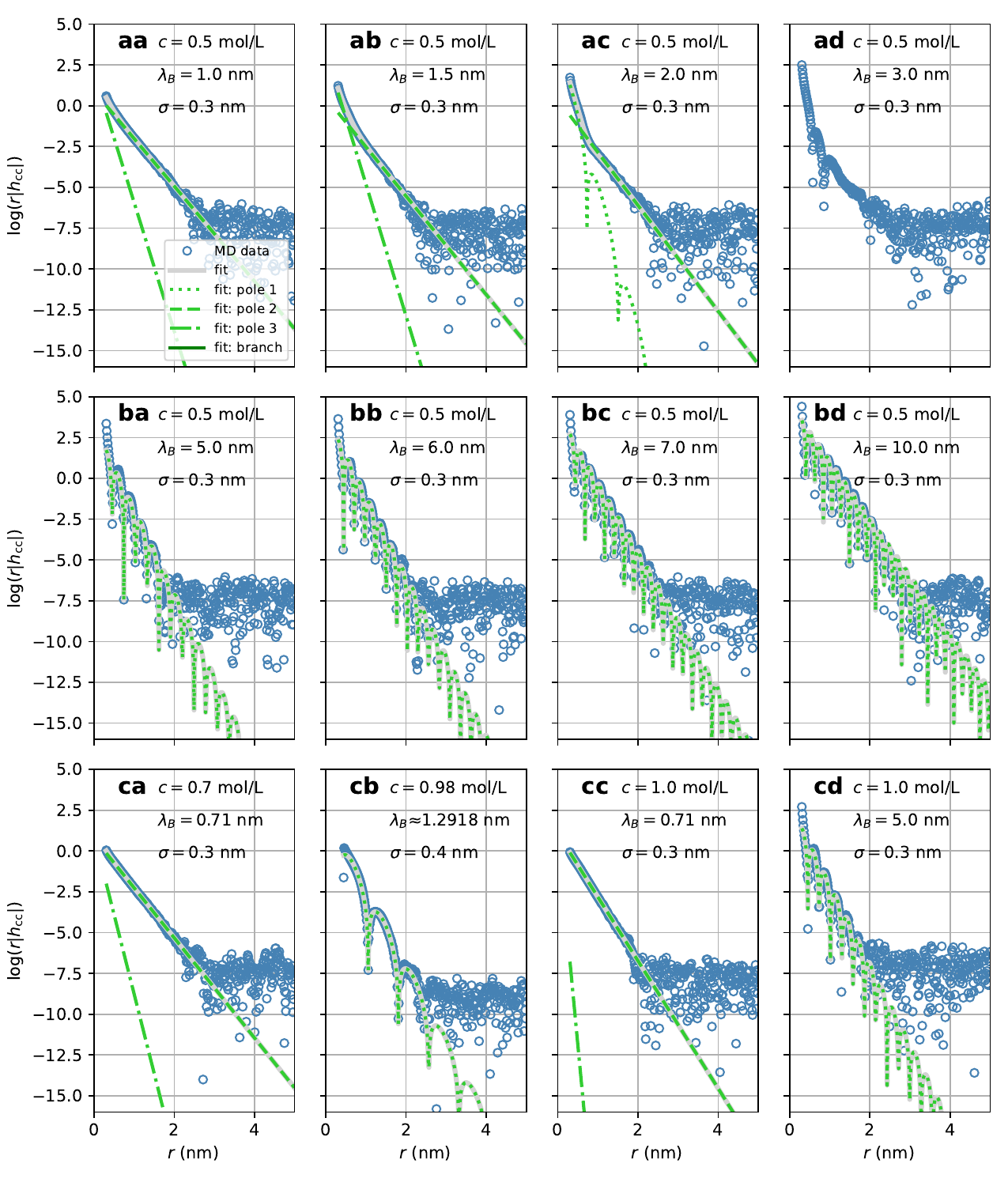}
\caption{Charge-correlation functions $h_{\textrm{cc}}(r)$ as explained in 
\cref{Sfig:corrfits01}, but for other parameter sets, as indicated and listed in \cref{Stab:sim-details}. }
\label{Sfig:corrfits08}
\end{figure*}

\begin{figure*}[t]
\centering
\includegraphics[width=\linewidth]{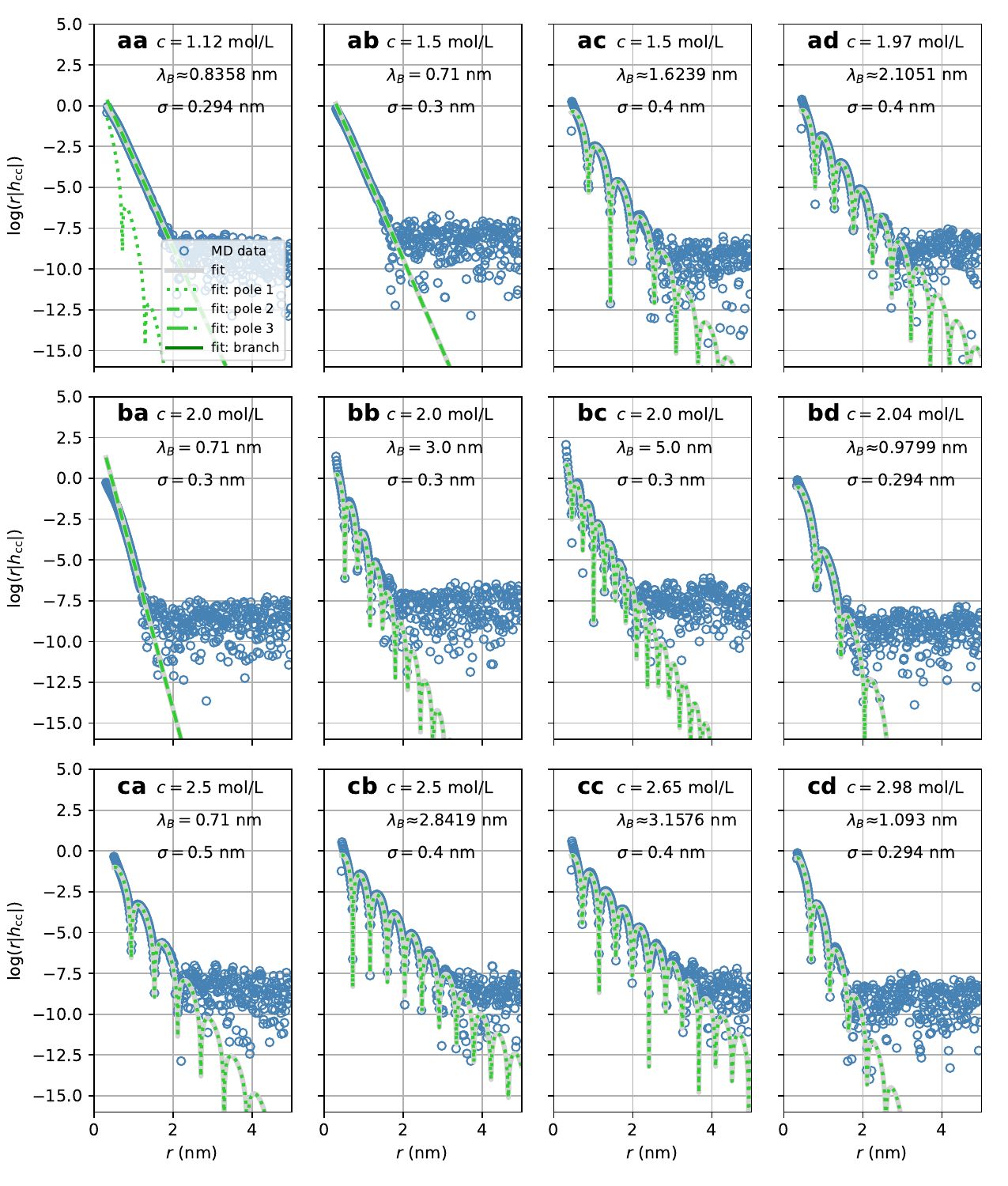}
\caption{Charge-correlation functions $h_{\textrm{cc}}(r)$ as explained in 
\cref{Sfig:corrfits01}, but for other parameter sets, as indicated and listed in \cref{Stab:sim-details}. }
\label{Sfig:corrfits09}
\end{figure*}

\begin{figure*}[t]
\centering
\includegraphics[width=\linewidth]{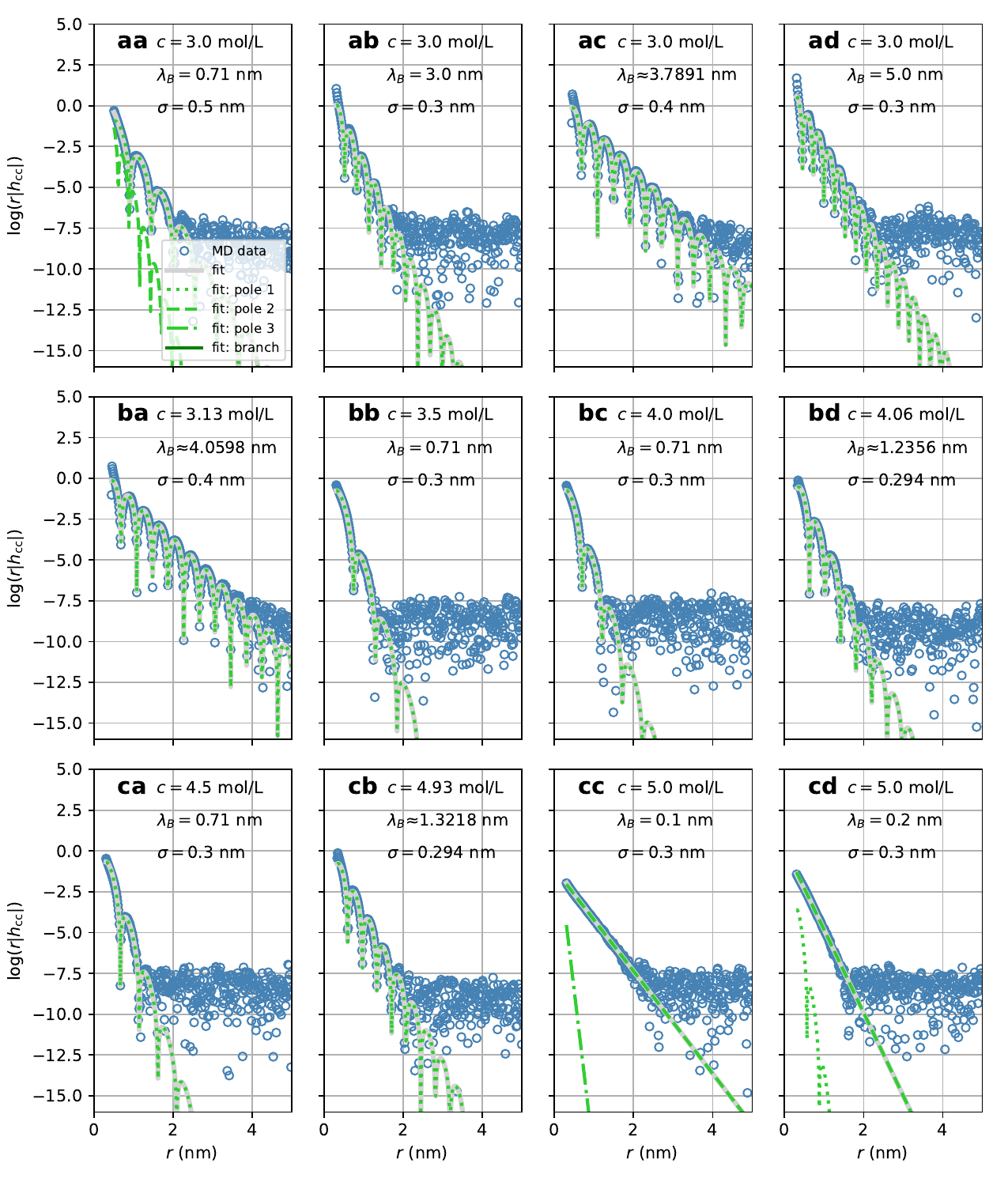}
\caption{Charge-correlation functions $h_{\textrm{cc}}(r)$ as explained in 
\cref{Sfig:corrfits01}, but for other parameter sets, as indicated and listed in \cref{Stab:sim-details}. }
\label{Sfig:corrfits10}
\end{figure*}

\begin{figure*}[t]
\centering
\includegraphics[width=\linewidth]{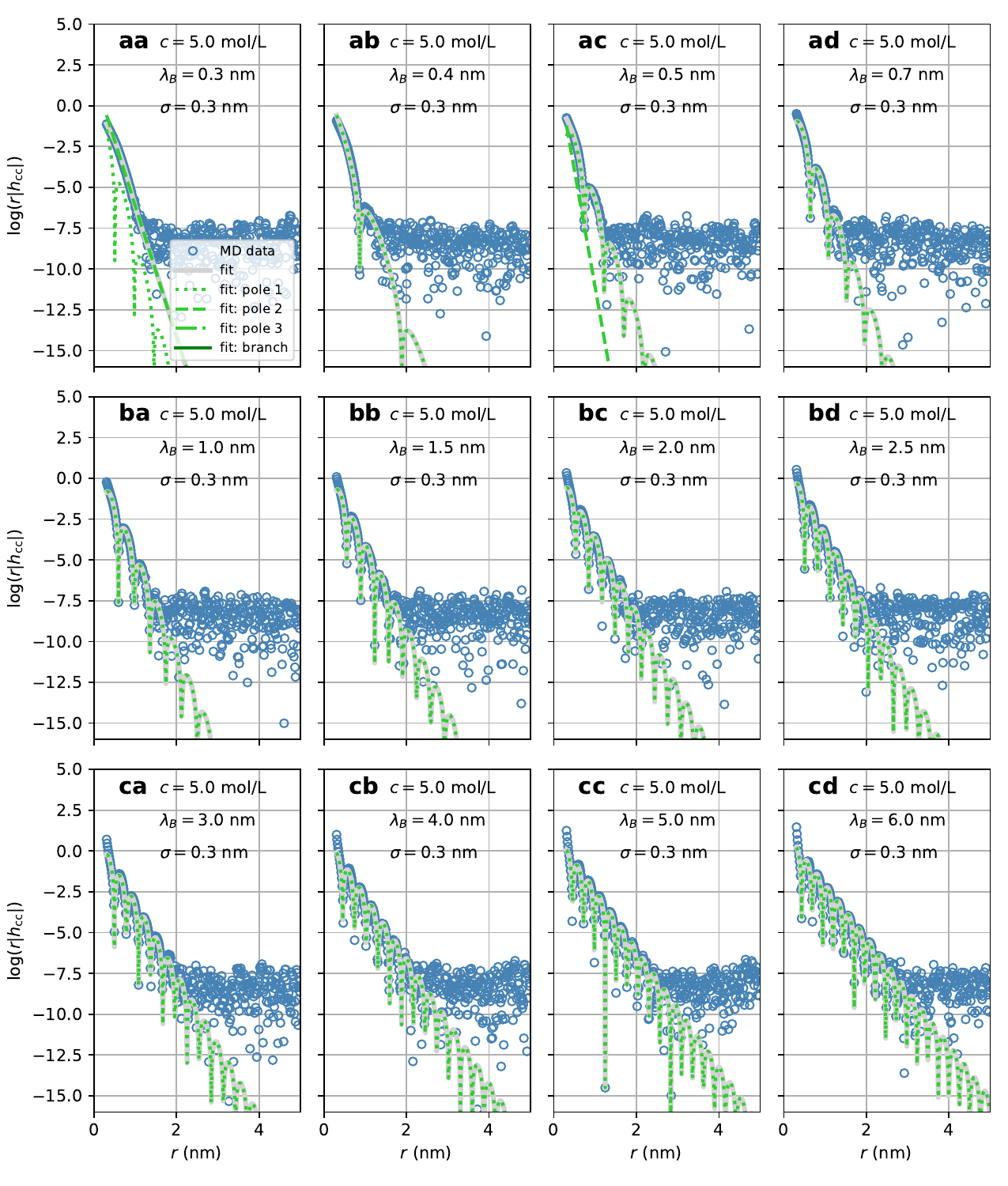}
\caption{Charge-correlation functions $h_{\textrm{cc}}(r)$ as explained in 
\cref{Sfig:corrfits01}, but for other parameter sets, as indicated and listed in \cref{Stab:sim-details}. }
\label{Sfig:corrfits11}
\end{figure*}

\begin{figure*}[t]
\centering
\includegraphics[width=\linewidth]{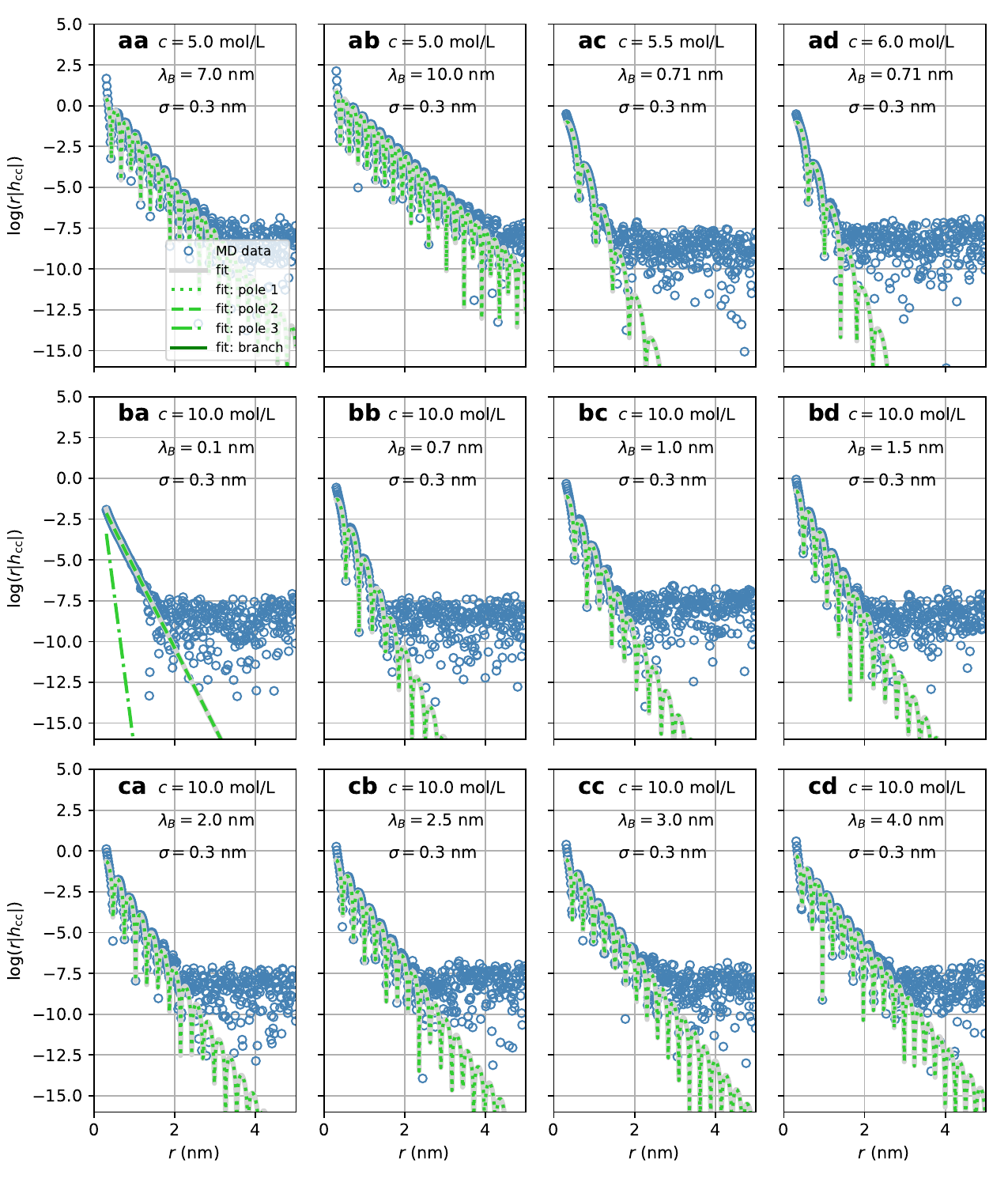}
\caption{Charge-correlation functions $h_{\textrm{cc}}(r)$ as explained in 
\cref{Sfig:corrfits01}, but for other parameter sets, as indicated and listed in \cref{Stab:sim-details}. }
\label{Sfig:corrfits12}
\end{figure*}

\begin{figure*}[t]
\centering
\includegraphics[width=\linewidth]{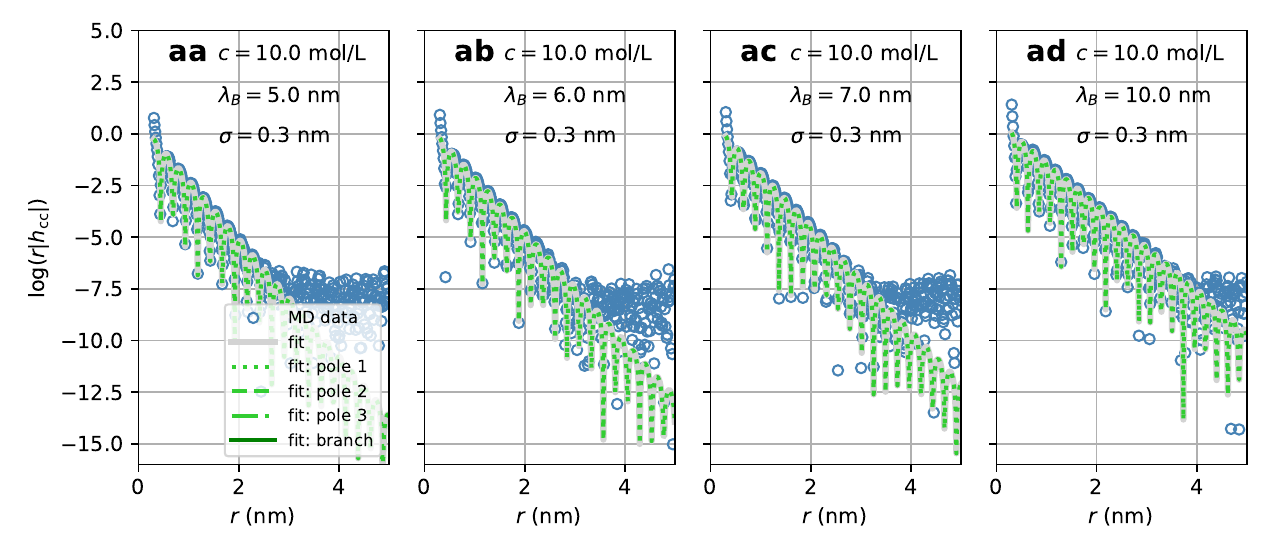}
\caption{Charge-correlation functions $h_{\textrm{cc}}(r)$ as explained in 
\cref{Sfig:corrfits01}, but for other parameter sets, as indicated and listed in \cref{Stab:sim-details}. }
\label{Sfig:corrfits13}
\end{figure*}

To extract decay lengths from our MD simulations, we perform fits to the sampled 
charge-correlation functions. This method is sufficient for our purposes, but we acknowledge 
the existence of more advanced methods of extracting asymptotic decay lengths 
\cite{Sulander_jcp114_2001,Sgonzalez-mozuelos_jcp139_2013}. 
Depending on the complexity of the simulation result, we use up to three pole contributions 
$P_i$ from \cref{Seq:pole-contribution} and up to one branch contribution $B_i$ from 
\cref{Seq:branch-contribution} to fit the sampled charge-correlation function. 
The range which we can subject to the fit is limited by numerical noise and is chosen manually 
because the point at which the correlation has decayed to the level of statistical noise varies heavily with the system parameters. 
All sampled charge-correlation functions, the resulting fits, and their decomposition into contributions $P_i(r)$ or $B_i(r)$ 
are shown in \cref{Sfig:corrfits01,Sfig:corrfits02,Sfig:corrfits03,Sfig:corrfits04,Sfig:corrfits05,Sfig:corrfits06,Sfig:corrfits07,Sfig:corrfits08,Sfig:corrfits09,Sfig:corrfits10,Sfig:corrfits11,Sfig:corrfits12,Sfig:corrfits13}. The corresponding fitting parameters are 
listed in \cref{Stab:sim-details} alongside the reference to the figure and panel displaying the fit. 
Each plot of the figures shows $\log(r|h_{\textrm{cc}}(r)|)$. 
In this representation the slope of the graph encodes the decay length of the corresponding
contribution $P_i$ in \cref{Seq:pole-contribution}. 

In \cref{Sfig:corrfits05}(bd) we exemplify the procedure. 
The fit slightly deviates from the signal in the region where the decay changes 
between different regimes, i.e., 
from pole 1 to pole 2. Close to particle contact (at $r=\sigma$), the argument of the exponential of each contribution 
is too small to scale-separate the longest decay length from sub-dominant contributions of other poles which also causes deviations. 
At larger separations the signal of our charge-correlation function becomes 
noisy hinging at the numerical limitations of our approach.

\section{Additional results from Density Functional Theory}

To complement our results, we additionally calculate charge-correlation functions
using classical density functional theory (DFT). 
In DFT, the thermodynamic grand potential is expressed as a functional of the one-body number 
density profiles $\rho_{\nu}(\vec{r})$ of all species $\nu$ that depend on the spatial 
position $\vec{r}$ in the total volume $V$, respectively. 
The equilibrium density profiles follow from a minimization principle $\delta\Omega/\delta\rho_{\nu}=0$ for 
this functional $\Omega(T,V,\mu_{\nu};[\{\rho_{\nu}\}])$ 
at given temperature $T$, volume $V$, and chemical potentials $\mu_{\nu}$ 
in external potentials $V_{\textrm{ext},\nu}(\vec{r})$ \cite{Sevans_ap28_1979,Shansen_book_2013}. 
For technical reasons, the functional typically is split into 
a free energy contribution $F_{\textrm{id}}$ from an ideal gas reference system, 
a potential contribution 
$\sum_{\nu}\int_V \rho_{\nu}(\vec{r})(V_{\textrm{ext},\nu}(\vec{r})-\mu_{\nu})\textrm{d}\vec{r}$ from 
Legendre transforms and the external potentials, and an excess free energy contribution 
from all pair interactions between the particles \cite{Shaertel_jpcm29_2017}. 
The latter contribution is 
typically not known exactly and, for this reason, we used four different approaches, 
namely the MSAc functional \cite{Sroth_jpcm28_2016,Scats_jcp154_2021}, 
the $\delta$- and $\theta$-functional \cite{Sbueltmann_jpcm24_2022}, 
and the electrostatic mean-field functional \cite{Shaertel_jpcm29_2017}. 
All these electrostatic contributions were supplemented by a hard-sphere 
contribution from fundamental measure theory \cite{Srosenfeld_prl63_1989,Sroth_jpcm22_2010}, 
the so-called White-Bear mark II version \cite{Shansen-goos_jpcm18_2006}. 

The MSAc functional and the $\delta$- and $\theta$-functional are well-performing 
approaches to describe correlation functions in the RPM 
in the framework of DFT \cite{Scats_jcp154_2021,Sbueltmann_jpcm24_2022}. 
While the former has been used in a recent study on 
underscreening \cite{Scats_jcp154_2021}, here we present results on underscreening 
for the other two functionals. The mean-field functional is included for 
completeness but it consistently predicts the Debye length and thus cannot account for any kind of underscreening. 
For the MSAc functional, the decay length was extracted from the far-field 
of the density profiles in a system with a planar charged hard wall  
which yields the same asymptotic as correlations in bulk \cite{Scats_jcp154_2021}. 
For the other three functionals, we directly calculate the charge-correlation functions 
using the so-called Percus trick \cite{Spercus_prl8_1962,Sfrisch_book_1964}. 
In the Percus trick, one particle in the bulk system is fixed by an external 
field and the one-body distribution in this field yields the corresponding 
bulk pair-distribution functions.

\begin{figure}[t]
\centering
\includegraphics[width=\linewidth]{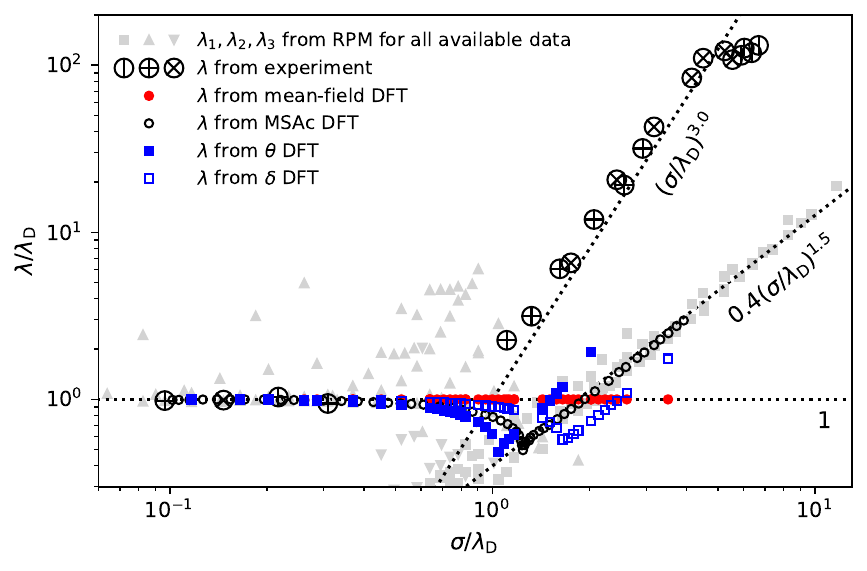}
\caption{Decay lengths $\lambda$ in relation to the Debye screening length $\lambda_{\textrm{D}}$ 
as a function of the ratio $\sigma/\lambda_{\textrm{D}}$. Gray symbols present the decay lengths 
from all decay contributions that we extracted from our MD simulations. Respective data is 
listed in \cref{Stab:sim-details}. 
Large black circles represent data from experiments on an ionic liquid (|), 
NaCl in water (+), and [C4C1Pyrr][NTf2] in propylene carbonate (X) \cite{Ssmith_jpcl7_2016}.
Dotted lines show power laws as noted. 
In addition, we re-show decay lengths 
from DFT calculations using the MSAc functional \cite{Scats_jcp154_2021} 
that were calculated  
at fixed Bjerrum length 
$\lambda_{\textrm{B}}=0.73$ nm and $\sigma=0.5$ nm 
as a function of the concentration. These results are compared against 
calculations using three other DFT approaches at various concentrations and Bjerrum 
lengths for $\sigma=0.3$ nm. }
\label{Sfig:decay-lengths}
\end{figure}

In \cref{Sfig:decay-lengths} the decay lengths of all our extracted 
decay contributions are depicted as gray symbols. They clearly show 
anomalous underscreening, as discussed in detail in the main article. 
Further, experimental results from \cite{Ssmith_jpcl7_2016} are shown (data listed in \cref{Stab:exp-decay}). 
In addition to the MSAc DFT data from \cite{Scats_jcp154_2021}, we 
present results that we calculated via the $\delta$- and $\theta$-functional 
DFT approaches \cite{Sbueltmann_jpcm24_2022} and via an electrostatic mean-field 
functional, as explained above. 
The mean-field approach always predicts the Debye length as the screening 
length and, hence, cannot predict underscreening at all. 
While the position of the Kirkwood transition from monotonic to oscillatory decay 
\cite{Skirkwood_jcp7_1939,SleoteDeCarvalho_mp83_1994} 
slightly differs between all three 
DFT approaches (compare discussion on the Kirkwood point in 
\cite{Scats_jcp154_2021}), the data from all three approaches 
share an increase of the decay 
length that follows roughly the same power law corresponding to 
regular underscreening. None of the three approaches predicts anomalous 
underscreening, even though the resolution of the correlation functions 
is much better than that of our MD results. 

We stress that the data from the MSAc functional are taken from \cite{Scats_jcp154_2021} 
and have been calculated by only varying the concentration, while we varied both 
concentration and Bjerrum length for the three other functional approaches. 
All these variations lead evidently to the same curve underlining that within the theory the ratio of screening length and Debye 
length depends exclusively on $\sigma/\lambda_{\textrm{D}}$. 
Thus, the regular underscreening predicted by DFT calculations appears universal in 
the representation of \cref{Sfig:decay-lengths} in contrast to our simulation 
results for anomalous underscreening.

\begin{table*}[ht]
\caption{\label{Stab:exp-decay} Parameter sets as reported for experimental measurements 
on an ionic liquid, NaCl in water, and [C4C1Pyrr][NTf2] in propylene carbonate from \cite{Ssmith_jpcl7_2016}. 
We list the reported concentration $c$, particle diameter $\sigma$, relative permittivity $\epsilon$, 
and the respective measured decay length $\lambda$. 
The experiments are performed at $T=294$ K. 
The Bjerrum length follows via $\lambda_{\textrm{B}}=e^2/(4\pi \epsilon_0\epsilon k_{\textrm{B}}T)$, 
with vacuum permittivity $\epsilon_0$ and elementary charge $e$. }
\centering
\begin{tabular}{|l|l|l|l|l|}
\hline
 & $c$ (mol/L) & $\lambda$ (nm) & $\sigma$ (nm) & $\epsilon$ \\
\hline
ionic liquid & 3.31 & 8.4 & 0.4  & 12.5 \\
\textit{from Table S1 and S4 in \cite{Ssmith_jpcl7_2016}} & 3.91 & 7.1 & 0.38 & 12.0 \\
& 6.54 & 6.3 & 0.32 & 12.9 \\
& 5.34 & 6.6 & 0.34 & 15.2 \\
\hline
NaCl in water & 0.01 & 3.0 & 0.294 & 80.0 \\
\textit{from Table S2 and S4(2nd) in \cite{Ssmith_jpcl7_2016}} & 0.05 & 1.4 & 0.294 & 79.0 \\
& 0.10 & 0.9 & 0.294 & 78.0 \\
& 1.12 & 0.6 & 0.294 & 68.0 \\
& 1.48 & 0.7 & 0.294 & 63.0 \\
& 2.04 & 1.1 & 0.294 & 58.0 \\
& 2.98 & 1.7 & 0.294 & 52.0 \\
& 4.06 & 2.2 & 0.294 & 46.0 \\
& 4.93 & 3.2 & 0.294 & 43.0 \\
\hline
[C4C1Pyrr][NTf2] in propylene carbonate & 0.01 & 2.7 & 0.40 & 64.0 \\
\textit{from Table S3 and S5 in \cite{Ssmith_jpcl7_2016}} & 0.98 & 1.5 & 0.40 & 44.0 \\
& 1.50 & 3.4 & 0.40 & 35.0 \\
& 1.97 & 5.4 & 0.40 & 27.0 \\
& 2.50 & 8.1 & 0.40 & 20.0 \\
& 2.65 & 9.8 & 0.40 & 18.0 \\
& 3.00 & 9.3 & 0.40 & 15.0 \\
& 3.13 & 7.8 & 0.40 & 14.0 \\
\hline
\end{tabular}
\end{table*}

\include{SM_table}

\newpage
~~

\end{document}

%% file: SM_table.tex
\small
\begin{longtable*}{|l|l|l|l|l|l|l|lllll|l|l|}
\caption{\label{Stab:sim-details} Relevant parameters and extracted data for all MD simulations we performed. 
  Simulation parameters are the concentration $c$ of ions, the Bjerrum length $\lambda_{\textrm{B}}$, the ion diameter $\sigma$, the length $L_{\textrm{z}}$ of the simulation box, and the number $N_{\textrm{g}}$ of samples used for sampling a pair-distribution function. 
  For convenience we also list the ratio $\sigma/\lambda_{\textrm{D}}$ of ion diameter and Debye length. 
  For each simulation we further list the parameters for each extracted pole or branch point and its corresponding decay length $\lambda$. 
  The first column refers to the figure and panel where the respective pair charge-correlation function and the fit functions from poles and branch points are shown. } \\
 &  &  &  & & & & pole: & \multicolumn{4}{l|}{$r h_{\mathrm{cc}}(r)=a\cdot\exp(-b\cdot r)\cos(c\cdot r+d)$} & & \\
 &  &  &  & & & & branch: & \multicolumn{4}{l|}{$r h_{\mathrm{cc}}(r)=a\cdot\exp(-b\cdot r)\cos(c\cdot r+d)/r$} & & \\
Figure & $c$ (mol/L) & $\lambda_{\mathrm{B}}$ (nm) & $\sigma$ (nm) & $\sigma/\lambda_{\mathrm{D}}$ & $L$ (nm) & $N_{\textrm{g}}$ & & $a$ & $b$ & $c$ & $d$ & $\lambda$ & $\lambda/\lambda_{\mathrm{D}}$ \\
\hline
\endfirsthead 
\caption{Continuation from previous page}\\
Figure & $c$ (mol/L) & $\lambda_{\mathrm{B}}$ (nm) & $\sigma$ (nm) & $\sigma/\lambda_{\mathrm{D}}$ & $L$ (nm) & $N_{\textrm{g}}$ & & $a$ (1) & $b$ (1/nm) & $c$ (1/nm) & $d$ (1) & $\lambda$ (nm) & $\lambda/\lambda_{\mathrm{D}}$ \\
\hline
\endhead 
\multicolumn{14}{r}{\textit{Table continues on the next page}}\\
\endfoot
\hline
\endlastfoot
8(aa) & $0.001$ & $0.71$ & $0.5$ & $0.05(1)$ & 120 & 74148 & pole 2: & 1.478(8) & 0.109(9) & 0.0 & 0.0 & 9.092(7) & 0.94(2) \\
 &  &  &  & & & & pole 3: & 4.199(6) & 4.782(2) & 0.0 & 0.0 & 0.209(1) & 0.02(1) \\
8(ab) & $0.001$ & $3.0$ & $0.3$ & $0.06(3)$ & 200 & 1268532 & pole 2: & 347.069(0) & 4.635(6) & 0.0 & 0.0 & 0.215(7) & 0.04(5) \\
 &  &  &  & & & & pole 3: & 3.721(7) & 0.196(8) & 0.0 & 0.0 & 5.081(0) & 1.08(2) \\
8(ac) & $0.001$ & $5.0$ & $0.3$ & $0.08(2)$ & 200 & 1230957 & pole 2: & 0.4 & 0.113 & 0.0 & 0.0 & 8.849(5) & 2.43(4) \\
8(ad) & $0.002$ & $0.71$ & $0.5$ & $0.07(3)$ & 120 & 64128 & pole 2: & 1.469(9) & 0.162(1) & 0.0 & 0.0 & 6.168(6) & 0.90(4) \\
 &  &  &  & & & & pole 3: & 4.286(8) & 4.933(6) & 0.0 & 0.0 & 0.202(6) & 0.02(9) \\
8(ba) & $0.002$ & $3.0$ & $0.3$ & $0.09(0)$ & 120 & 824145 & pole 2: & 444.217(1) & 5.771(9) & 0.0 & 0.0 & 0.173(2) & 0.05(2) \\
 &  &  &  & & & & pole 3: & 3.598(3) & 0.282(1) & 0.0 & 0.0 & 3.543(8) & 1.06(7) \\
8(bb) & $0.003$ & $0.71$ & $0.5$ & $0.08(9)$ & 120 & 54108 & pole 2: & 1.396(6) & 0.167(5) & 0.0 & 0.0 & 5.968(1) & 1.07(1) \\
 &  &  &  & & & & pole 3: & 1.919(0) & 3.345(8) & 0.0 & 0.0 & 0.298(8) & 0.05(3) \\
8(bc) & $0.003$ & $3.0$ & $0.3$ & $0.11(0)$ & 120 & 706331 & pole 2: & 3.264(4) & 0.327(6) & 0.0 & 0.0 & 3.052(2) & 1.12(6) \\
 &  &  &  & & & & pole 3: & 453.512(1) & 6.202(1) & 0.0 & 0.0 & 0.161(2) & 0.05(9) \\
8(bd) & $0.005$ & $0.71$ & $0.5$ & $0.11(5)$ & 120 & 19038 & pole 2: & 1.365(7) & 0.208(0) & 0.0 & 0.0 & 4.806(8) & 1.11(4) \\
 &  &  &  & & & & pole 3: & 1.405(4) & 2.944(7) & 0.0 & 0.0 & 0.339(5) & 0.07(8) \\
8(ca) & $0.005$ & $3.0$ & $0.3$ & $0.14(2)$ & 120 & 886269 & pole 2: & 1.778(0) & 0.357(9) & 0.0 & 0.0 & 2.793(8) & 1.33(0) \\
 &  &  &  & & & & pole 3: & 310.124(8) & 5.680(8) & 0.0 & 0.0 & 0.176(0) & 0.08(3) \\
8(cb) & $0.005$ & $5.0$ & $0.3$ & $0.18(4)$ & 120 & 1098693 & pole 2: & 0.057(6) & 0.194(7) & 0.0 & 0.0 & 5.133(4) & 3.15(7) \\
8(cc) & $0.01$ & $0.71$ & $0.5$ & $0.16(3)$ & 30 & 19038 & pole 2: & 1.493(3) & 0.337(1) & 0.0 & 0.0 & 2.966(3) & 0.97(2) \\
 &  &  &  & & & & pole 3: & 4.588(2) & 5.621(4) & 0.0 & 0.0 & 0.177(8) & 0.05(8) \\
8(cd) & $0.01$ & $\approx0.8881$ & $0.4$ & $0.14(6)$ & 60 & 184368 & pole 2: & 1.841(2) & 0.376(9) & 0.0 & 0.0 & 2.652(9) & 0.97(2) \\
 &  &  &  & & & & pole 3: & 6.381(9) & 4.784(4) & 0.0 & 0.0 & 0.209(0) & 0.07(6) \\
\hline
9(aa) & $0.01$ & $3.0$ & $0.3$ & $0.20(2)$ & 120 & 391281 & pole 2: & 1.111(7) & 0.446(5) & 0.0 & 0.0 & 2.239(6) & 1.50(8) \\
9(ab) & $0.01$ & $5.0$ & $0.3$ & $0.26(0)$ & 120 & 693384 & pole 2: & 0.022(5) & 0.175 & 0.0 & 0.0 & 5.714(2) & 4.97(0) \\
9(ac) & $0.02$ & $0.71$ & $0.5$ & $0.23(1)$ & 60 & 19038 & pole 2: & 1.518(6) & 0.472(5) & 0.0 & 0.0 & 2.116(2) & 0.98(0) \\
 &  &  &  & & & & pole 3: & 5.565(0) & 6.223(8) & 0.0 & 0.0 & 0.160(6) & 0.07(4) \\
9(ad) & $0.02$ & $3.0$ & $0.3$ & $0.28(5)$ & 60 & 480960 & pole 2: & 0.761(8) & 0.582(1) & 0.0 & 0.0 & 1.717(7) & 1.63(6) \\
9(ba) & $0.04$ & $0.71$ & $0.5$ & $0.32(7)$ & 60 & 16533 & pole 2: & 1.528(2) & 0.669(7) & 0.0 & 0.0 & 1.493(0) & 0.97(8) \\
 &  &  &  & & & & pole 3: & 6.420(8) & 6.495(1) & 0.0 & 0.0 & 0.153(9) & 0.10(0) \\
9(bb) & $0.05$ & $0.1$ & $0.3$ & $0.08(2)$ & 60 & 32064 & pole 2: & 0.205(6) & 0.282(4) & 0.0 & 0.0 & 3.540(2) & 0.97(3) \\
9(bc) & $0.05$ & $0.2$ & $0.3$ & $0.11(6)$ & 60 & 32064 & pole 2: & 0.419(7) & 0.403(1) & 0.0 & 0.0 & 2.480(2) & 0.96(4) \\
9(bd) & $0.05$ & $0.3$ & $0.3$ & $0.14(2)$ & 60 & 32064 & pole 2: & 0.624(9) & 0.491(3) & 0.0 & 0.0 & 2.035(0) & 0.96(9) \\
 &  &  &  & & & & pole 3: & 1.838(3) & 11.088(8) & 0.0 & 0.0 & 0.090(1) & 0.04(2) \\
9(ca) & $0.05$ & $0.4$ & $0.3$ & $0.16(5)$ & 60 & 32064 & pole 2: & 0.830(1) & 0.560(5) & 0.0 & 0.0 & 1.784(1) & 0.98(1) \\
 &  &  &  & & & & pole 3: & 4.696(0) & 10.844(9) & 0.0 & 0.0 & 0.092(2) & 0.05(0) \\
9(cb) & $0.05$ & $0.5$ & $0.3$ & $0.18(4)$ & 60 & 32064 & pole 2: & 1.031(1) & 0.622(5) & 0.0 & 0.0 & 1.606(2) & 0.98(7) \\
 &  &  &  & & & & pole 3: & 3.169(0) & 7.855(0) & 0.0 & 0.0 & 0.127(3) & 0.07(8) \\
9(cc) & $0.05$ & $0.7$ & $0.3$ & $0.21(8)$ & 60 & 31563 & pole 2: & 1.409(2) & 0.730(0) & 0.0 & 0.0 & 1.369(7) & 0.99(6) \\
 &  &  &  & & & & pole 3: & 4.641(5) & 6.288(9) & 0.0 & 0.0 & 0.159(0) & 0.11(5) \\
9(cd) & $0.05$ & $1.0$ & $0.3$ & $0.26(0)$ & 60 & 32064 & pole 2: & 1.947(4) & 0.872(1) & 0.0 & 0.0 & 1.146(5) & 0.99(7) \\
 &  &  &  & & & & pole 3: & 17.966(4) & 7.130(2) & 0.0 & 0.0 & 0.140(2) & 0.12(1) \\
\hline
10(aa) & $0.05$ & $1.5$ & $0.3$ & $0.31(9)$ & 60 & 32064 & pole 2: & 2.371(1) & 1.018(3) & 0.0 & 0.0 & 0.981(9) & 1.04(6) \\
 &  &  &  & & & & pole 3: & 67.251(9) & 7.575(1) & 0.0 & 0.0 & 0.132(0) & 0.14(0) \\
10(ab) & $0.05$ & $2.0$ & $0.3$ & $0.36(9)$ & 60 & 32064 & pole 2: & 1.846(9) & 1.024(1) & 0.0 & 0.0 & 0.976(3) & 1.20(0) \\
10(ac) & $0.05$ & $2.5$ & $0.3$ & $0.41(2)$ & 60 & 32064 & pole 1: & 2010.106(4) & 9.264(4) & 1.504(9) & 1.073(4) & 0.107(9) & 0.14(8) \\
 &  &  &  & & & & pole 2: & 1.115(8) & 0.967(8) & 0.0 & 0.0 & 1.033(2) & 1.42(0) \\
 &  &  &  & & & & pole 3: & 241.112(3) & 6.689(4) & 0.0 & 0.0 & 0.149(4) & 0.20(5) \\
10(ad) & $0.05$ & $3.0$ & $0.3$ & $0.45(1)$ & 60 & 1088172 & pole 2: & 0.427(4) & 0.792(2) & 0.0 & 0.0 & 1.262(2) & 1.90(1) \\
 &  &  &  & & & & pole 3: & 3.105(2) & 3.223(3) & 0.0 & 0.0 & 0.310(2) & 0.46(7) \\
10(ba) & $0.05$ & $4.0$ & $0.3$ & $0.52(1)$ & 60 & 32064 & pole 1: & 499.825(9) & 9.397(1) & 10.046(1) & 2.449(1) & 0.106(4) & 0.18(5) \\
 &  &  &  & & & & pole 2: & 0.04 & 0.5 & 0.0 & 0.0 & 2.0 & 3.47(9) \\
 &  &  &  & & & & pole 3: & 1.448(5) & 3.021(9) & 0.0 & 0.0 & 0.330(9) & 0.57(5) \\
10(bb) & $0.05$ & $5.0$ & $0.3$ & $0.58(3)$ & 60 & 1181358 & pole 1: & 820.302(8) & 8.740(1) & 10.065(5) & -3.067(0) & 0.114(4) & 0.22(2) \\
 &  &  &  & & & & pole 2: & 0.015(4) & 0.608(0) & 0.0 & 0.0 & 1.644(7) & 3.19(8) \\
 &  &  &  & & & & pole 3: & 19.345(9) & 6.536(1) & 0.0 & 0.0 & 0.152(9) & 0.29(7) \\
10(bc) & $0.05$ & $6.0$ & $0.3$ & $0.63(9)$ & 60 & 32064 & pole 1: & 621.937(7) & 7.464(6) & 10.213(2) & 0.567(6) & 0.133(9) & 0.28(5) \\
10(bd) & $0.05$ & $7.0$ & $0.3$ & $0.69(0)$ & 60 & 32064 & pole 1: & 683.114(3) & 6.538(0) & 11.775(3) & 2.925(9) & 0.152(9) & 0.35(1) \\
10(ca) & $0.05$ & $10.0$ & $0.3$ & $0.82(5)$ & 60 & 32064 & pole 1: & 1978.496(8) & 5.131(7) & 14.247(8) & -1.066(8) & 0.194(8) & 0.53(5) \\
10(cb) & $0.06$ & $0.71$ & $0.5$ & $0.40(1)$ & 60 & 14028 & pole 2: & 1.716(4) & 0.862(1) & 0.0 & 0.0 & 1.159(8) & 0.93(1) \\
 &  &  &  & & & & pole 3: & 11.645(3) & 9.488(3) & 0.0 & 0.0 & 0.105(3) & 0.08(4) \\
10(cc) & $0.06$ & $3.0$ & $0.3$ & $0.49(5)$ & 60 & 975948 & pole 1: & 539.340(5) & 9.325(4) & 4.812(7) & -1.005(2) & 0.107(2) & 0.17(6) \\
 &  &  &  & & & & pole 2: & 0.438(1) & 0.884(0) & 0.0 & 0.0 & 1.131(2) & 1.86(6) \\
 &  &  &  & & & & pole 3: & 42.502(2) & 5.748(1) & 0.0 & 0.0 & 0.173(9) & 0.28(7) \\
10(cd) & $0.07$ & $3.0$ & $0.3$ & $0.53(4)$ & 60 & 924345 & pole 1: & 217.213(8) & 8.434(1) & 5.860(7) & -1.440(2) & 0.118(5) & 0.21(1) \\
 &  &  &  & & & & pole 2: & 0.416(5) & 0.944(8) & 0.0 & 0.0 & 1.058(3) & 1.88(6) \\
 &  &  &  & & & & pole 3: & 64.902(8) & 6.561(7) & 0.0 & 0.0 & 0.152(3) & 0.27(1) \\
\hline
11(aa) & $0.07$ & $5.0$ & $0.3$ & $0.69(0)$ & 60 & 1092681 & pole 1: & 120.001(6) & 6.770(4) & 10.509(5) & 2.771(2) & 0.147(6) & 0.33(9) \\
 &  &  &  & & & & pole 2: & 0.009(3) & 0.506(7) & 0.0 & 0.0 & 1.973(3) & 4.54(0) \\
 &  &  &  & & & & pole 3: & 15.925(5) & 5.712(1) & 0.0 & 0.0 & 0.175(0) & 0.40(2) \\
11(ab) & $0.08$ & $3.0$ & $0.3$ & $0.57(1)$ & 60 & 750999 & pole 1: & 25.132(6) & 7.721(9) & 12.012(9) & 0.138(7) & 0.129(5) & 0.24(6) \\
 &  &  &  & & & & pole 2: & 0.328(0) & 0.941(6) & 0.0 & 0.0 & 1.061(9) & 2.02(3) \\
 &  &  &  & & & & pole 3: & 1.657(9) & 3.160(0) & 0.0 & 0.0 & 0.316(4) & 0.60(3) \\
11(ac) & $0.08$ & $5.0$ & $0.3$ & $0.73(8)$ & 60 & 967544 & pole 1: & 120.002(1) & 6.704(3) & 10.602(0) & 2.699(7) & 0.149(1) & 0.36(6) \\
 &  &  &  & & & & pole 2: & 0.009(5) & 0.543(0) & 0.0 & 0.0 & 1.841(5) & 4.53(0) \\
 &  &  &  & & & & pole 3: & 10.665(3) & 5.436(3) & 0.0 & 0.0 & 0.183(9) & 0.45(2) \\
11(ad) & $0.09$ & $3.0$ & $0.3$ & $0.60(6)$ & 60 & 826650 & pole 1: & 25.467(1) & 7.650(6) & 11.537(8) & 0.598(9) & 0.130(7) & 0.26(4) \\
 &  &  &  & & & & pole 2: & 1.383(2) & 3.169(0) & 0.0 & 0.0 & 0.315(5) & 0.63(7) \\
 &  &  &  & & & & pole 3: & 0.314(7) & 0.993(5) & 0.0 & 0.0 & 1.006(4) & 2.03(4) \\
11(ba) & $0.09$ & $5.0$ & $0.3$ & $0.78(2)$ & 60 & 843181 & pole 1: & 120.0 & 6.799(9) & 10.5 & 2.783(1) & 0.147(0) & 0.38(3) \\
 &  &  &  & & & & pole 2: & 0.009(9) & 0.550(5) & 0.0 & 0.0 & 1.816(2) & 4.73(9) \\
 &  &  &  & & & & pole 3: & 10.659(2) & 5.567(4) & 0.0 & 0.0 & 0.179(6) & 0.46(8) \\
11(bb) & $0.1$ & $0.1$ & $0.3$ & $0.11(6)$ & 60 & 26049 & pole 2: & 0.207(2) & 0.357(5) & 0.0 & 0.0 & 2.796(8) & 1.08(7) \\
11(bc) & $0.1$ & $0.2$ & $0.3$ & $0.16(5)$ & 60 & 26052 & pole 2: & 0.425(2) & 0.523(4) & 0.0 & 0.0 & 1.910(2) & 1.05(0) \\
11(bd) & $0.1$ & $0.3$ & $0.3$ & $0.20(2)$ & 60 & 26052 & pole 2: & 0.653(5) & 0.652(7) & 0.0 & 0.0 & 1.532(0) & 1.03(2) \\
11(ca) & $0.1$ & $0.4$ & $0.3$ & $0.23(3)$ & 60 & 26052 & pole 2: & 0.867(5) & 0.746(3) & 0.0 & 0.0 & 1.339(9) & 1.04(2) \\
 &  &  &  & & & & pole 3: & 1.079(3) & 6.748(7) & 0.0 & 0.0 & 0.148(1) & 0.11(5) \\
11(cb) & $0.1$ & $0.5$ & $0.3$ & $0.26(0)$ & 60 & 26052 & pole 2: & 1.104(5) & 0.841(9) & 0.0 & 0.0 & 1.187(7) & 1.03(3) \\
 &  &  &  & & & & pole 3: & 5.940(7) & 9.686(8) & 0.0 & 0.0 & 0.103(2) & 0.08(9) \\
11(cc) & $0.1$ & $0.7$ & $0.3$ & $0.30(8)$ & 60 & 26052 & pole 2: & 1.456(5) & 0.969(4) & 0.0 & 0.0 & 1.031(5) & 1.06(1) \\
 &  &  &  & & & & pole 3: & 4.724(9) & 6.292(2) & 0.0 & 0.0 & 0.158(9) & 0.16(3) \\
11(cd) & $0.1$ & $1.0$ & $0.3$ & $0.36(9)$ & 60 & 26052 & pole 2: & 1.981(7) & 1.150(2) & 0.0 & 0.0 & 0.869(4) & 1.06(9) \\
 &  &  &  & & & & pole 3: & 18.312(4) & 7.270(5) & 0.0 & 0.0 & 0.137(5) & 0.16(9) \\
\hline
12(aa) & $0.1$ & $1.5$ & $0.3$ & $0.45(1)$ & 60 & 26052 & pole 2: & 2.320(4) & 1.334(2) & 0.0 & 0.0 & 0.749(4) & 1.12(9) \\
 &  &  &  & & & & pole 3: & 85.828(5) & 8.392(4) & 0.0 & 0.0 & 0.119(1) & 0.17(9) \\
12(ab) & $0.1$ & $2.0$ & $0.3$ & $0.52(1)$ & 60 & 26052 & pole 2: & 1.790(5) & 1.353(6) & 0.0 & 0.0 & 0.738(7) & 1.28(5) \\
 &  &  &  & & & & pole 3: & 259.287(1) & 9.358(5) & 0.0 & 0.0 & 0.106(8) & 0.18(5) \\
12(ac) & $0.1$ & $2.5$ & $0.3$ & $0.58(3)$ & 60 & 26052 & pole 2: & 0.917(7) & 1.235(7) & 0.0 & 0.0 & 0.809(2) & 1.57(3) \\
 &  &  &  & & & & pole 3: & 1528.0 & 12.4 & 0.0 & 0.0 & 0.080(6) & 0.15(6) \\
12(ad) & $0.1$ & $3.0$ & $0.3$ & $0.63(9)$ & 60 & 578649 & pole 2: & 0.332(7) & 1.065(4) & 0.0 & 0.0 & 0.938(5) & 1.99(9) \\
12(ba) & $0.1$ & $3.5$ & $0.3$ & $0.69(0)$ & 60 & 969432 & pole 2: & 0.138(5) & 0.913(1) & 0.0 & 0.0 & 1.095(0) & 2.51(9) \\
12(bb) & $0.1$ & $4.0$ & $0.3$ & $0.73(8)$ & 60 & 995982 & pole 1: & 131.943(7) & 7.734(2) & 10.617(2) & 2.215(8) & 0.129(2) & 0.31(8) \\
 &  &  &  & & & & pole 2: & 0.055(8) & 0.784(8) & 0.0 & 0.0 & 1.274(2) & 3.13(4) \\
12(bc) & $0.1$ & $4.5$ & $0.3$ & $0.78(2)$ & 60 & 1510004 & pole 1: & 350.0 & 8.259(7) & 9.682(9) & -2.953(5) & 0.121(0) & 0.31(5) \\
 &  &  &  & & & & pole 2: & 0.021(7) & 0.667(9) & 0.0 & 0.0 & 1.497(2) & 3.90(6) \\
12(bd) & $0.1$ & $5.0$ & $0.3$ & $0.82(5)$ & 60 & 1546754 & pole 1: & 303.511(5) & 7.509(2) & 9.8 & -2.641(5) & 0.133(1) & 0.36(6) \\
 &  &  &  & & & & pole 2: & 0.011(5) & 0.65 & 0.0 & 0.0 & 1.538(4) & 4.23(1) \\
12(ca) & $0.1$ & $5.5$ & $0.3$ & $0.86(5)$ & 60 & 1514513 & pole 1: & 143.104(5) & 6.304 & 11.164(6) & -3.453(7) & 0.158(6) & 0.45(7) \\
 &  &  &  & & & & pole 2: & 0.005(7) & 0.587(9) & 0.0 & 0.0 & 1.700(9) & 4.90(6) \\
12(cb) & $0.1$ & $6.0$ & $0.3$ & $0.90(3)$ & 60 & 973938 & pole 1: & 253.214(9) & 6.384(5) & 10.825(6) & -2.970(4) & 0.156(6) & 0.47(1) \\
 &  &  &  & & & & pole 2: & 0.003 & 0.5 & 0.0 & 0.0 & 2.0 & 6.02(5) \\
12(cc) & $0.1$ & $7.0$ & $0.3$ & $0.97(6)$ & 60 & 945883 & pole 1: & 324.065(9) & 5.704(6) & 12.104(1) & -0.435(6) & 0.175(2) & 0.57(0) \\
12(cd) & $0.1$ & $10.0$ & $0.3$ & $1.16(6)$ & 60 & 926150 & pole 1: & 1687.226(1) & 4.897(3) & 14.378(1) & 1.978(5) & 0.204(1) & 0.79(4) \\
\hline
13(aa) & $0.15$ & $0.71$ & $0.5$ & $0.63(4)$ & 60 & 14028 & pole 2: & 1.951(4) & 1.374(7) & 0.0 & 0.0 & 0.727(4) & 0.92(3) \\
 &  &  &  & & & & pole 3: & 8.781(4) & 10.792(1) & 0.0 & 0.0 & 0.092(6) & 0.11(7) \\
13(ab) & $0.2$ & $0.1$ & $0.3$ & $0.16(5)$ & 60 & 12024 & pole 2: & 0.203(3) & 0.553(8) & 0.0 & 0.0 & 1.805(6) & 0.99(3) \\
13(ac) & $0.2$ & $0.2$ & $0.3$ & $0.23(3)$ & 60 & 12024 & pole 2: & 0.424(6) & 0.797(9) & 0.0 & 0.0 & 1.253(1) & 0.97(4) \\
 &  &  &  & & & & pole 3: & 6.137(6) & 19.551(6) & 0.0 & 0.0 & 0.051(1) & 0.03(9) \\
13(ad) & $0.2$ & $0.3$ & $0.3$ & $0.28(5)$ & 60 & 12024 & pole 2: & 0.648(1) & 0.975(7) & 0.0 & 0.0 & 1.024(8) & 0.97(6) \\
 &  &  &  & & & & pole 3: & 6.474(7) & 15.734(2) & 0.0 & 0.0 & 0.063(5) & 0.06(0) \\
13(ba) & $0.2$ & $0.4$ & $0.3$ & $0.33(0)$ & 60 & 11523 & pole 2: & 0.851(1) & 1.112(4) & 0.0 & 0.0 & 0.898(8) & 0.98(8) \\
 &  &  &  & & & & pole 3: & 0.725(5) & 6.646(8) & 0.0 & 0.0 & 0.150(4) & 0.16(5) \\
13(bb) & $0.2$ & $0.5$ & $0.3$ & $0.36(9)$ & 60 & 12024 & pole 2: & 1.097(9) & 1.260(5) & 0.0 & 0.0 & 0.793(2) & 0.97(5) \\
 &  &  &  & & & & pole 3: & 6.382(7) & 11.197(1) & 0.0 & 0.0 & 0.089(3) & 0.10(9) \\
13(bc) & $0.2$ & $0.7$ & $0.3$ & $0.43(6)$ & 60 & 12024 & pole 2: & 1.650(0) & 1.530(3) & 0.0 & 0.0 & 0.653(4) & 0.95(0) \\
 &  &  &  & & & & pole 3: & 12.455(4) & 11.325(4) & 0.0 & 0.0 & 0.088(2) & 0.12(8) \\
13(bd) & $0.2$ & $1.0$ & $0.3$ & $0.52(1)$ & 60 & 12024 & pole 2: & 2.035(7) & 1.753(2) & 0.0 & 0.0 & 0.570(3) & 0.99(2) \\
 &  &  &  & & & & pole 3: & 14.891(3) & 8.355(3) & 0.0 & 0.0 & 0.119(6) & 0.20(8) \\
13(ca) & $0.2$ & $1.5$ & $0.3$ & $0.63(9)$ & 60 & 9519 & pole 2: & 2.193(3) & 2.000(9) & 0.0 & 0.0 & 0.499(7) & 1.06(4) \\
 &  &  &  & & & & pole 3: & 51.449(1) & 8.825(8) & 0.0 & 0.0 & 0.113(3) & 0.24(1) \\
13(cb) & $0.2$ & $2.0$ & $0.3$ & $0.73(8)$ & 60 & 12024 & pole 2: & 1.450(7) & 1.971(0) & 0.0 & 0.0 & 0.507(3) & 1.24(8) \\
 &  &  &  & & & & pole 3: & 146.587(5) & 9.884(5) & 0.0 & 0.0 & 0.101(1) & 0.24(8) \\
13(cc) & $0.2$ & $2.5$ & $0.3$ & $0.82(5)$ & 60 & 12024 & pole 2: & 0.697(2) & 1.765(7) & 0.0 & 0.0 & 0.566(3) & 1.55(7) \\
 &  &  &  & & & & pole 3: & 592.819(5) & 12.562(7) & 0.0 & 0.0 & 0.079(6) & 0.21(8) \\
13(cd) & $0.2$ & $3.0$ & $0.3$ & $0.90(3)$ & 60 & 12024 & pole 1: & 450.682(2) & 10.787(9) & 8.051(5) & -3.111(1) & 0.092(6) & 0.27(9) \\
 &  &  &  & & & & branch: & 0.199(6) & 0.978(3) & 0.0 & 0.0 & 1.022(1) & 3.07(9) \\
\hline
14(aa) & $0.2$ & $4.0$ & $0.3$ & $1.04(3)$ & 60 & 12024 & pole 1: & 1192.387(2) & 10.512(7) & 10.451(3) & 2.429(2) & 0.095(1) & 0.33(0) \\
 &  &  &  & & & & pole 2: & 0.068(0) & 1.227(2) & 0.0 & 0.0 & 0.814(8) & 2.83(4) \\
14(ab) & $0.2$ & $5.0$ & $0.3$ & $1.16(6)$ & 60 & 12024 & pole 1: & 118.493(7) & 6.515(5) & 10.280(7) & -3.012(1) & 0.153(4) & 0.59(6) \\
14(ac) & $0.2$ & $6.0$ & $0.3$ & $1.27(8)$ & 60 & 12024 & pole 1: & 111.605(4) & 5.559(8) & 11.310(5) & -0.137(7) & 0.179(8) & 0.76(6) \\
14(ad) & $0.2$ & $7.0$ & $0.3$ & $1.38(0)$ & 60 & 12024 & pole 1: & 130.494(6) & 4.814(8) & 12.426(4) & -0.545(0) & 0.207(6) & 0.95(5) \\
14(ba) & $0.2$ & $10.0$ & $0.3$ & $1.65(0)$ & 60 & 9823 & pole 1: & 470.323(5) & 4.299(5) & 14.446(2) & 1.960(6) & 0.232(5) & 1.27(9) \\
14(bb) & $0.3$ & $0.71$ & $0.3$ & $0.53(8)$ & 30 & 16533 & pole 2: & 1.604(7) & 1.849(7) & 0.0 & 0.0 & 0.540(6) & 0.97(0) \\
 &  &  &  & & & & pole 3: & 3.285(9) & 7.395(8) & 0.0 & 0.0 & 0.135(2) & 0.24(2) \\
14(bc) & $0.5$ & $0.1$ & $0.3$ & $0.26(0)$ & 60 & 3808 & pole 2: & 0.213(8) & 0.890(0) & 0.0 & 0.0 & 1.123(5) & 0.97(7) \\
 &  &  &  & & & & pole 3: & 0.301(7) & 24.915(5) & 0.0 & 0.0 & 0.040(1) & 0.03(4) \\
14(bd) & $0.5$ & $0.2$ & $0.3$ & $0.36(9)$ & 60 & 3307 & pole 2: & 0.457(5) & 1.282(4) & 0.0 & 0.0 & 0.779(7) & 0.95(9) \\
 &  &  &  & & & & pole 3: & 4.057(7) & 25.408(5) & 0.0 & 0.0 & 0.039(3) & 0.04(8) \\
14(ca) & $0.5$ & $0.3$ & $0.3$ & $0.45(1)$ & 60 & 3307 & pole 2: & 0.698(1) & 1.561(3) & 0.0 & 0.0 & 0.640(4) & 0.96(4) \\
 &  &  &  & & & & pole 3: & 7.395(6) & 17.103(5) & 0.0 & 0.0 & 0.058(4) & 0.08(8) \\
14(cb) & $0.5$ & $0.4$ & $0.3$ & $0.52(1)$ & 60 & 3808 & pole 2: & 0.967(8) & 1.819(8) & 0.0 & 0.0 & 0.549(4) & 0.95(5) \\
 &  &  &  & & & & pole 3: & 2.102(7) & 11.397(8) & 0.0 & 0.0 & 0.087(7) & 0.15(2) \\
14(cc) & $0.5$ & $0.5$ & $0.3$ & $0.58(3)$ & 60 & 3307 & pole 2: & 1.272(9) & 2.061(8) & 0.0 & 0.0 & 0.485(0) & 0.94(3) \\
 &  &  &  & & & & pole 3: & 3.166(3) & 11.415(1) & 0.0 & 0.0 & 0.087(6) & 0.17(0) \\
14(cd) & $0.5$ & $0.7$ & $0.3$ & $0.69(0)$ & 60 & 3307 & pole 2: & 1.892(3) & 2.481(1) & 0.0 & 0.0 & 0.403(0) & 0.92(7) \\
 &  &  &  & & & & pole 3: & 6.725(5) & 10.919(3) & 0.0 & 0.0 & 0.091(5) & 0.21(0) \\
\hline
15(aa) & $0.5$ & $1.0$ & $0.3$ & $0.82(5)$ & 60 & 3307 & pole 2: & 2.469(6) & 2.915(7) & 0.0 & 0.0 & 0.342(9) & 0.94(3) \\
 &  &  &  & & & & pole 3: & 7.180(4) & 7.895(0) & 0.0 & 0.0 & 0.126(6) & 0.34(8) \\
15(ab) & $0.5$ & $1.5$ & $0.3$ & $1.01(0)$ & 60 & 3150 & pole 2: & 1.643(2) & 3.019(5) & 0.0 & 0.0 & 0.331(1) & 1.11(5) \\
 &  &  &  & & & & pole 3: & 26.571(0) & 8.087(0) & 0.0 & 0.0 & 0.123(6) & 0.41(6) \\
15(ac) & $0.5$ & $2.0$ & $0.3$ & $1.16(6)$ & 60 & 3142 & pole 1: & 51.277(8) & 8.657(4) & 4.037(8) & -1.387(5) & 0.115(5) & 0.44(9) \\
 &  &  &  & & & & pole 2: & 1.482(8) & 3.247(3) & 0.0 & 0.0 & 0.307(9) & 1.19(7) \\
15(ad) & $0.5$ & $3.0$ & $0.3$ & $1.42(9)$ & 60 & 3170 &  & & & & & &  \\
15(ba) & $0.5$ & $5.0$ & $0.3$ & $1.84(5)$ & 60 & 3307 & pole 1: & 27.267(1) & 5.131(2) & 10.776(9) & -0.162(3) & 0.194(8) & 1.19(8) \\
15(bb) & $0.5$ & $6.0$ & $0.3$ & $2.02(1)$ & 60 & 3307 & pole 1: & 48.144(1) & 4.902(3) & 11.867(8) & 2.585(3) & 0.203(9) & 1.37(4) \\
15(bc) & $0.5$ & $7.0$ & $0.3$ & $2.18(3)$ & 60 & 3307 & pole 1: & 54.675(0) & 4.206(7) & 12.909(0) & -0.932(5) & 0.237(7) & 1.72(9) \\
15(bd) & $0.5$ & $10.0$ & $0.3$ & $2.60(9)$ & 60 & 3117 & pole 1: & 101.867(8) & 3.511(4) & 14.494(0) & -1.220(7) & 0.284(7) & 2.47(6) \\
15(ca) & $0.7$ & $0.71$ & $0.3$ & $0.82(2)$ & 30 & 15531 & pole 2: & 2.241(4) & 3.064(3) & 0.0 & 0.0 & 0.326(3) & 0.89(4) \\
 &  &  &  & & & & pole 3: & 2.672(9) & 9.736(3) & 0.0 & 0.0 & 0.102(7) & 0.28(1) \\
15(cb) & $0.98$ & $\approx1.2918$ & $0.4$ & $1.75(0)$ & 30 & 160821 & pole 1: & 10.701(1) & 4.625(9) & 4.164(2) & 0.322(3) & 0.216(1) & 0.94(6) \\
15(cc) & $1.0$ & $0.71$ & $0.3$ & $0.98(3)$ & 30 & 16533 & pole 2: & 2.996(8) & 3.912(5) & 0.0 & 0.0 & 0.255(5) & 0.83(7) \\
 &  &  &  & & & & pole 3: & 3.273(1) & 26.018(4) & 0.0 & 0.0 & 0.038(4) & 0.12(5) \\
15(cd) & $1.0$ & $5.0$ & $0.3$ & $2.60(9)$ & 20 & 8016 & pole 1: & 17.862(9) & 4.876(8) & 11.124(8) & -0.348(2) & 0.205(0) & 1.78(3) \\
\hline
16(aa) & $1.12$ & $\approx0.8358$ & $0.294$ & $1.10(6)$ & 30 & 163829 & pole 1: & 16.714(8) & 10.259(5) & 5.415(1) & 0.833(5) & 0.097(4) & 0.36(6) \\
 &  &  &  & & & & pole 2: & 8.671(2) & 5.428(9) & 0.0 & 0.0 & 0.184(1) & 0.69(3) \\
16(ab) & $1.5$ & $0.71$ & $0.3$ & $1.20(4)$ & 30 & 14529 & pole 2: & 6.261(2) & 5.601(0) & 0.0 & 0.0 & 0.178(5) & 0.71(6) \\
16(ac) & $1.5$ & $\approx1.6239$ & $0.4$ & $2.42(8)$ & 30 & 108266 & pole 1: & 5.838(1) & 3.883(4) & 5.664(8) & -0.254(0) & 0.257(5) & 1.56(3) \\
16(ad) & $1.97$ & $\approx2.1051$ & $0.4$ & $3.16(8)$ & 30 & 51051 & pole 1: & 4.062(1) & 3.320(0) & 6.466(0) & -0.398(4) & 0.301(1) & 2.38(5) \\
16(ba) & $2.0$ & $0.71$ & $0.3$ & $1.39(0)$ & 30 & 13527 & pole 2: & 57.646(2) & 9.088(6) & 0.0 & 0.0 & 0.110(0) & 0.50(9) \\
16(bb) & $2.0$ & $3.0$ & $0.3$ & $2.85(8)$ & 20 & 8016 & pole 1: & 9.014(8) & 5.675(8) & 9.864(1) & 2.618(6) & 0.176(1) & 1.67(8) \\
16(bc) & $2.0$ & $5.0$ & $0.3$ & $3.69(0)$ & 20 & 8016 & pole 1: & 9.760(4) & 4.491(6) & 11.538(5) & 2.525(0) & 0.222(6) & 2.73(8) \\
16(bd) & $2.04$ & $\approx0.9799$ & $0.294$ & $1.61(6)$ & 30 & 53185 & pole 1: & 9.705(2) & 6.490(3) & 5.209(8) & -2.788(4) & 0.154(0) & 0.84(7) \\
16(ca) & $2.5$ & $0.71$ & $0.5$ & $2.59(1)$ & 30 & 9519 & pole 1: & 3.680(7) & 3.934(2) & 5.344(8) & 2.825(9) & 0.254(1) & 1.31(7) \\
16(cb) & $2.5$ & $\approx2.8419$ & $0.4$ & $4.14(7)$ & 30 & 11438 & pole 1: & 3.020(6) & 2.782(4) & 7.190(7) & 2.596(3) & 0.359(3) & 3.72(6) \\
16(cc) & $2.65$ & $\approx3.1576$ & $0.4$ & $4.50(0)$ & 30 & 7237 & pole 1: & 2.684(2) & 2.599(0) & 7.463(9) & 2.459(9) & 0.384(7) & 4.32(9) \\
16(cd) & $2.98$ & $\approx1.093$ & $0.294$ & $2.06(3)$ & 30 & 18652 & pole 1: & 6.898(5) & 5.952(1) & 6.647(0) & -3.029(1) & 0.168(0) & 1.17(9) \\
\hline
17(aa) & $3.0$ & $0.71$ & $0.5$ & $2.83(8)$ & 30 & 6513 & pole 1: & 3.697(3) & 3.921(9) & 5.705(1) & -0.415(3) & 0.254(9) & 1.44(7) \\
 &  &  &  & & & & pole 2: & 17.258(8) & 8.140(8) & 11.543(6) & -2.362(4) & 0.122(8) & 0.69(7) \\
17(ab) & $3.0$ & $3.0$ & $0.3$ & $3.50(0)$ & 20 & 5511 & pole 1: & 5.765(3) & 5.088(8) & 10.170(8) & 2.608(1) & 0.196(5) & 2.29(3) \\
17(ac) & $3.0$ & $\approx3.7891$ & $0.4$ & $5.24(5)$ & 30 & 6432 & pole 1: & 2.770(9) & 2.416(0) & 7.760(8) & 2.508(8) & 0.413(8) & 5.42(7) \\
17(ad) & $3.0$ & $5.0$ & $0.3$ & $4.51(9)$ & 20 & 5511 & pole 1: & 6.431(9) & 4.170(0) & 11.682(1) & 2.501(6) & 0.239(8) & 3.61(2) \\
17(ba) & $3.13$ & $\approx4.0598$ & $0.4$ & $5.54(6)$ & 30 & 13244 & pole 1: & 2.598(4) & 2.312(5) & 7.937(2) & -0.732(8) & 0.432(4) & 5.99(5) \\
17(bb) & $3.5$ & $0.71$ & $0.3$ & $1.83(9)$ & 30 & 5511 & pole 1: & 7.251(4) & 7.142(1) & 5.724(8) & 0.426(8) & 0.140(0) & 0.85(8) \\
17(bc) & $4.0$ & $0.71$ & $0.3$ & $1.96(6)$ & 30 & 4509 & pole 1: & 6.103(9) & 6.951(7) & 6.150(9) & -2.745(5) & 0.143(8) & 0.94(2) \\
17(bd) & $4.06$ & $\approx1.2356$ & $0.294$ & $2.56(1)$ & 30 & 9330 & pole 1: & 4.861(3) & 5.361(8) & 7.994(8) & 2.812(3) & 0.186(5) & 1.62(4) \\
17(ca) & $4.5$ & $0.71$ & $0.3$ & $2.08(5)$ & 30 & 2004 & pole 1: & 6.588(7) & 7.101(9) & 6.623(0) & -2.853(8) & 0.140(8) & 0.97(8) \\
17(cb) & $4.93$ & $\approx1.3218$ & $0.294$ & $2.91(9)$ & 30 & 4824 & pole 1: & 2.690(8) & 4.611(2) & 8.503(7) & -0.358(6) & 0.216(8) & 2.15(3) \\
17(cc) & $5.0$ & $0.1$ & $0.3$ & $0.82(5)$ & 20 & 5010 & pole 2: & 0.340(4) & 3.138(2) & 0.0 & 0.0 & 0.318(6) & 0.87(6) \\
 &  &  &  & & & & pole 3: & 5.507(9) & 20.387(0) & 0.0 & 0.0 & 0.049(0) & 0.13(4) \\
17(cd) & $5.0$ & $0.2$ & $0.3$ & $1.16(6)$ & 20 & 5010 & pole 1: & 5.816(9) & 15.164(8) & 10.075(5) & -1.008(3) & 0.065(9) & 0.25(6) \\
 &  &  &  & & & & pole 2: & 1.299(6) & 5.084(8) & 0.0 & 0.0 & 0.196(6) & 0.76(4) \\
\hline
18(aa) & $5.0$ & $0.3$ & $0.3$ & $1.42(9)$ & 20 & 5010 & pole 1: & 4.343(4) & 9.471(2) & 6.624(0) & 1.361(0) & 0.105(5) & 0.50(2) \\
 &  &  &  & & & & pole 2: & 6.500(6) & 8.006(4) & 0.0 & 0.0 & 0.124(8) & 0.59(5) \\
18(ab) & $5.0$ & $0.4$ & $0.3$ & $1.65(0)$ & 20 & 5010 & pole 1: & 5.034(9) & 7.163(5) & 3.079(3) & 2.041(6) & 0.139(5) & 0.76(7) \\
18(ac) & $5.0$ & $0.5$ & $0.3$ & $1.84(5)$ & 20 & 5010 & pole 1: & 3.996(0) & 7.090(5) & 6.518(5) & 3.065(6) & 0.141(0) & 0.86(7) \\
 &  &  &  & & & & pole 2: & 23.485(4) & 14.274(7) & 0.0 & 0.0 & 0.070(0) & 0.43(0) \\
18(ad) & $5.0$ & $0.7$ & $0.3$ & $2.18(3)$ & 20 & 4509 & pole 1: & 4.342(8) & 6.569(6) & 7.163(3) & 0.083(7) & 0.152(2) & 1.10(7) \\
18(ba) & $5.0$ & $1.0$ & $0.3$ & $2.60(9)$ & 20 & 4509 & pole 1: & 3.763(2) & 5.887(6) & 8.222(1) & -0.173(9) & 0.169(8) & 1.47(7) \\
18(bb) & $5.0$ & $1.5$ & $0.3$ & $3.19(5)$ & 20 & 4008 & pole 1: & 3.055(0) & 5.094(4) & 9.263(6) & -0.408(2) & 0.196(2) & 2.09(0) \\
18(bc) & $5.0$ & $2.0$ & $0.3$ & $3.69(0)$ & 20 & 4008 & pole 1: & 2.728(7) & 4.587(9) & 9.848(9) & 2.665(7) & 0.217(9) & 2.68(1) \\
18(bd) & $5.0$ & $2.5$ & $0.3$ & $4.12(5)$ & 20 & 4008 & pole 1: & 3.259(1) & 4.537(1) & 10.229(3) & 2.663(7) & 0.220(4) & 3.03(1) \\
18(ca) & $5.0$ & $3.0$ & $0.3$ & $4.51(9)$ & 20 & 4008 & pole 1: & 3.312(0) & 4.331(0) & 10.697(6) & 2.579(1) & 0.230(8) & 3.47(8) \\
18(cb) & $5.0$ & $4.0$ & $0.3$ & $5.21(8)$ & 20 & 4008 & pole 1: & 2.906(3) & 3.813(5) & 11.010(5) & -0.282(5) & 0.262(2) & 4.56(1) \\
18(cc) & $5.0$ & $5.0$ & $0.3$ & $5.83(4)$ & 20 & 4008 & pole 1: & 3.213(2) & 3.591(1) & 11.874(1) & 2.510(2) & 0.278(4) & 5.41(5) \\
18(cd) & $5.0$ & $6.0$ & $0.3$ & $6.39(1)$ & 20 & 4008 & pole 1: & 3.372(7) & 3.264(4) & 12.340(6) & -0.630(6) & 0.306(3) & 6.52(6) \\
\hline
19(aa) & $5.0$ & $7.0$ & $0.3$ & $6.90(3)$ & 20 & 4008 & pole 1: & 3.914(1) & 3.032(5) & 12.985(8) & 2.291(0) & 0.329(7) & 7.58(8) \\
19(ab) & $5.0$ & $10.0$ & $0.3$ & $8.25(1)$ & 20 & 4008 & pole 1: & 5.019(7) & 2.315(1) & 14.376(0) & 1.966(6) & 0.431(9) & 11.88(0) \\
19(ac) & $5.5$ & $0.71$ & $0.3$ & $2.30(5)$ & 30 & 2505 & pole 1: & 3.879(8) & 6.401(5) & 7.600(8) & -0.039(8) & 0.156(2) & 1.20(0) \\
19(ad) & $6.0$ & $0.71$ & $0.3$ & $2.40(8)$ & 30 & 2505 & pole 1: & 3.658(9) & 6.348(0) & 7.893(3) & -0.105(9) & 0.157(5) & 1.26(4) \\
19(ba) & $10.0$ & $0.1$ & $0.3$ & $1.16(6)$ & 20 & 1503 & pole 2: & 0.526(6) & 4.884(2) & 0.0 & 0.0 & 0.204(7) & 0.79(6) \\
 &  &  &  & & & & pole 3: & 11.281(9) & 18.996(3) & 0.0 & 0.0 & 0.052(6) & 0.20(4) \\
19(bb) & $10.0$ & $0.7$ & $0.3$ & $3.08(7)$ & 20 & 1503 & pole 1: & 1.894(5) & 5.533(0) & 9.585(7) & 2.662(6) & 0.180(7) & 1.85(9) \\
19(bc) & $10.0$ & $1.0$ & $0.3$ & $3.69(0)$ & 20 & 501 & pole 1: & 1.675(0) & 4.850(5) & 10.244(6) & -0.546(1) & 0.206(1) & 2.53(5) \\
19(bd) & $10.0$ & $1.5$ & $0.3$ & $4.51(9)$ & 20 & 1503 & pole 1: & 1.911(6) & 4.366(8) & 10.887(3) & 2.525(0) & 0.228(9) & 3.44(9) \\
19(ca) & $10.0$ & $2.0$ & $0.3$ & $5.21(8)$ & 20 & 1002 & pole 1: & 1.847(5) & 3.922(3) & 11.293(7) & -0.640(3) & 0.254(9) & 4.43(4) \\
19(cb) & $10.0$ & $2.5$ & $0.3$ & $5.83(4)$ & 20 & 1002 & pole 1: & 1.792(3) & 3.605(8) & 11.621(9) & 2.464(5) & 0.277(3) & 5.39(3) \\
19(cc) & $10.0$ & $3.0$ & $0.3$ & $6.39(1)$ & 20 & 1002 & pole 1: & 1.714(0) & 3.345(0) & 11.931(7) & 2.370(3) & 0.298(9) & 6.36(8) \\
19(cd) & $10.0$ & $4.0$ & $0.3$ & $7.38(0)$ & 20 & 1002 & pole 1: & 1.974(1) & 3.123(6) & 12.417(4) & 2.264(0) & 0.320(1) & 7.87(5) \\
\hline
20(aa) & $10.0$ & $5.0$ & $0.3$ & $8.25(1)$ & 20 & 501 & pole 1: & 1.861(9) & 2.848(0) & 12.764(2) & 2.217(5) & 0.351(1) & 9.65(7) \\
20(ab) & $10.0$ & $6.0$ & $0.3$ & $9.03(8)$ & 20 & 1002 & pole 1: & 1.822(6) & 2.659(5) & 13.024(4) & 2.210(8) & 0.376(0) & 11.32(8) \\
20(ac) & $10.0$ & $7.0$ & $0.3$ & $9.76(3)$ & 20 & 1002 & pole 1: & 1.939(5) & 2.542(5) & 13.316(2) & 2.185(3) & 0.393(3) & 12.79(9) \\
20(ad) & $10.0$ & $10.0$ & $0.3$ & $11.66(9)$ & 20 & 1002 & pole 1: & 1.974(9) & 2.050(1) & 14.182(6) & 2.059(9) & 0.487(7) & 18.97(2) \\
\hline
\end{longtable*}